\documentclass[notitlepage,twocolumn,prl,tightenlines,superscriptaddress,showpacs,floatfix]{revtex4-1} 

\usepackage{amsmath,amssymb,amsfonts,latexsym}
\usepackage{ulem}
\usepackage[colorlinks=true, linkcolor=blue, citecolor=blue]{hyperref}
\usepackage{bbm}
\usepackage{graphicx}
\usepackage{epsfig}
\usepackage{color}
\usepackage[activate=normal]{pdfcprot}  
\usepackage{bm} 
\usepackage{psfrag}
\usepackage{txfonts,pxfonts,wasysym}
\usepackage{pifont}
\usepackage[caption = false]{subfig}

\newcommand{\be}{\begin{equation}}
\newcommand{\ee}{\end{equation}}
\newcommand{\ben}{\begin{equation*}}
\newcommand{\een}{\end{equation*}}
\newcommand{\ba}{\begin{eqnarray}}
\newcommand{\ea}{\end{eqnarray}}

\newcommand{\gulli}{Gulliver Lab, UMR CNRS 7083, PSL Research University, ESPCI Paris 10 rue Vauquelin, 75005 Paris, France}
\newcommand{\lptmc}{Sorbonne Universit\'{e}, CNRS UMR 7600, Laboratoire de Physique Th\'{e}orique de la Matière Condens\'{e}e, LPTMC, 4 place Jussieu, Couloir 12-13, 5ème \'{e}tage, 75252 Paris Cedex 05, France.}
\newcommand{\iuf}{Institut Universitaire de France, 1 rue Descartes, 75231 Paris Cedex 05, France}
\newcommand{\lpthe}{Sorbonne Universit\'{e}, CNRS UMR 7589, Laboratoire de Physique Th\'{e}orique et Hautes Energies, LPTHE, 4 place Jussieu, Couloir 13-14, 5ème \'{e}tage, 75252 Paris Cedex 05, France}

\begin{document}

\title{Velocity and Speed Correlations in Hamiltonian Flocks} 

\author{Mathias Casiulis}
\affiliation{\lptmc}
\author{Marco Tarzia}
\affiliation{\lptmc}
\affiliation{\iuf}
\author{Leticia F. Cugliandolo}
\affiliation{\lpthe}
\affiliation{\iuf}
\author{Olivier Dauchot}
\affiliation{\gulli}

\date{\today}

\begin{abstract}
We study 
a $2d$ Hamiltonian fluid made of particles carrying spins coupled to their velocities.  At low temperatures and intermediate densities, this conservative system exhibits phase coexistence between a collectively moving droplet and a still gas.  The particle displacements within the droplet have remarkably similar correlations to those 
of birds flocks. The center of mass behaves as  
an effective self-propelled particle, driven by the droplet's total magnetization. The conservation of a generalized angular momentum leads to rigid rotations, opposite to the 
fluctuations of the magnetization orientation that,
however small, are responsible for the shape and scaling of the correlations.
\end{abstract}

\maketitle

Flocking, the formation of compact groups of collectively moving individuals, is a hallmark of animal group behavior. 
This phenomenon has long attracted the interest of biologists~\cite{Breder1954,Aoki1982,Badgerow1988,Huth1992,Krause2002,Tunstrom2013,Rosenthal2015},
and motivated a large number of theoretical studies, connecting microscopic Vicsek like models~\cite{Vicsek1995,Gregoire2004,Vicsek2012} to continuous theories of the Toner-Tu type~\cite{Toner95,Tu1998,Toner2005,Peshkov2014,Marchetti2013,Yang2015}. On the experimental side, most quantitative data were obtained in artificial systems~\cite{Schaller2010,Deseigne2010,Bricard2013,Geyer2018}. 
One noticeable exception is the large scale observational and data analysis effort conducted by the Starflag project~\cite{Cavagna2010,Cavagna2013a,Attanasi2014a,Hemelrijk2014,Attanasi2015,Cavagna2018,Cavagna2019}:
the individual three-dimensional trajectories of a few thousand birds in compact flocks were obtained and studied. In particular, the starling flocks display 
correlations for the bird speeds and velocities that have been described as long-ranged and scale-free~\cite{Cavagna2010,Cavagna2019}. 

The study of collective motion in vibrated polar grains~\cite{Deseigne2012,Weber2013} showed that considering the particles' headings and their velocities as distinct, but coupled, degrees of freedom was instrumental to model the experimental observations~\cite{Deseigne2010,Scholz2018}. It was soon noticed that, 
under such coupling, the existence of collective motion {\it in equilibrium} cannot be 
ruled out by standard arguments. A Hamiltonian model of particles carrying ferro\--magnetically coupled spins, that also interact with their own velocities, was then proposed~\cite{Bore2016}. 
Such model exhibits collectively moving polar ground states at temperature $T=0$  (in finite size systems)~\cite{Bore2016}, 
\begin{figure}[t!]
\centering
\includegraphics[width = 0.96\columnwidth]{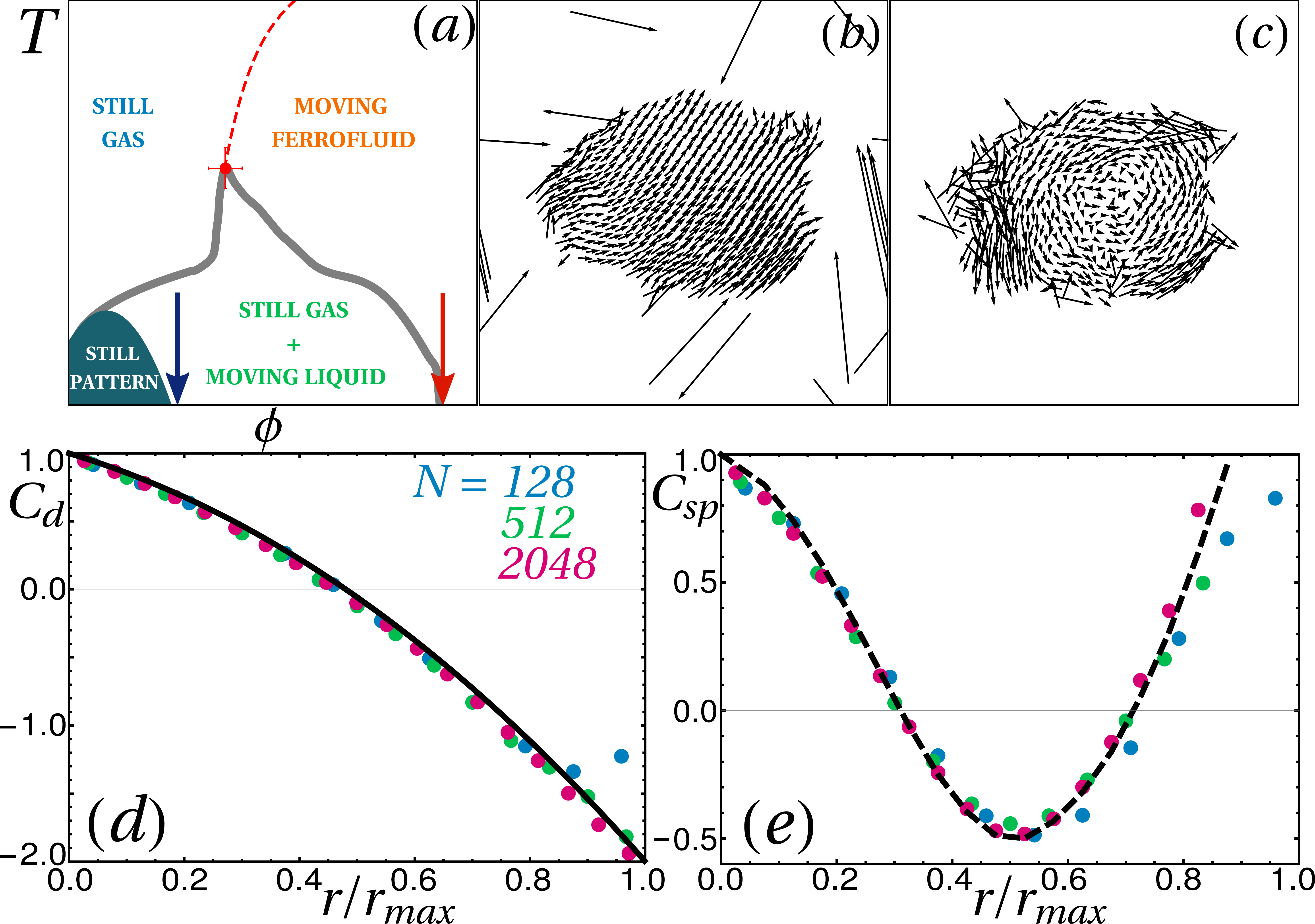}
\vspace{-2.5mm}
\caption{ {\bf Hamiltonian flocks.} 
$(a)$ Sketch of the phase diagram in the ($\phi, T$) plane, for small $K$ and $N$. 
The red and blue arrows indicate the densities studied here, $\phi = 0.55$ and $\phi = 0.14$.
$(b)-(c)$ Snapshots of the displacements and their fluctuations within a flock ($N = 512, K = 0.03, T = 10^{-2}$). 
$(d)-(e)$ Spatial correlations $C_d$ (resp. $C_{sp}$) of the displacement (resp. the speed) fluctuations, against distance normalized by the typical droplet size, for three system sizes ($T = 10^{-2}$). The black lines are our theoretical predictions.}
\vspace{-6mm}
\label{fig:HamFlock}
\end{figure}
and a rich phase diagram at finite $T$~\cite{Casiulis2019b}, Fig.~\ref{fig:HamFlock}(a), with  coexistence 
between a magnetized moving droplet and a disordered still gas, see Figs.~\ref{fig:HamFlock}(b)-(c) and \textit{Movie1.mp4} in the SI.

In this Letter, we show that the velocities and speeds of the particles inside the moving droplet exhibit correlations that are very similar to those observed in bird flocks, Figs.~\ref{fig:HamFlock}(d)-(e).
Analyzing the evolution in the homogeneous polar and moving droplet phases, we show that the dynamics of the center of mass are the ones of an effective self-propelled particle, driven by the droplet magnetization. The spontaneous fluctuations of the magnetization, together with the conservation of a generalized angular momentum, induce a
rigid body rotation of the droplet which, 
we show,  is responsible for the form and scaling of the displacement correlations. We conclude by discussing the relevance of our findings for real bird flocks.

Our model, first introduced in~\cite{Bore2016}, consists in a system of $N$ particles, in a 2d periodic square box of linear size $L_0$, interacting through an isotropic short-range repulsive potential $U(r)=4(1 - r)^4 \Theta(1 - r)$, with $\Theta$ the Heaviside step function. Each particle carries a unit planar vector, or spin, ${\bm{s}}_i$, with continuous orientation. The spins are coupled ferro-magnetically through an isotropic short-range coupling $J(r)=(1 - r)^2 \Theta(1 - r)$. This ferromagnetic interaction induces an effective attraction at low temperature, $T$,  resulting in an effective hard radius $r_0 = 1/4$ at $T=0$. We 
define the packing fraction $\phi \equiv \pi N r_0^2 / L_0^2$.
 The key ingredient is the introduction of a self-alignment between the spin and the velocity of each particle, through a coupling constant $K$. Starting from the Lagrangian formulation (see the SI), we obtain:
\begin{align}
   \mathcal{H} &\!=\!\!\sum\limits_{i = 1}^{N}\!\left( \frac{\bm{p}_i^2}{2}\!+\! 
                \frac{\omega_i^2}{2} - K \bm{p}_i\! \cdot \!\bm{s}_i \right)
  \!+\! \frac{1}{2} \sum\limits_{k \neq i}\!\left[ U(r_{ik}) \!-\!  J(r_{ik}) \, \bm{s}_i\!\cdot\!\bm{s}_k \right]
\label{eq:Ham}
\end{align}
where ${\bm{r}_i}$ are the positions of the particles, $r_{ik}= |{\bm r}_i - {\bm r}_k|$, ${\bm{s}_i}$ their spins parametrized by an angle $\theta_i$, $\omega_i = \dot{\theta_i}$ the momenta associated to these angles, and $\bm{p}_i = \dot{\bm{r}}_i + K \bm{s}_i$ the linear momenta associated 
to the positions.

The dynamics conserve the total energy $\mathcal{H}$, the total linear momentum $\bm{P} = \sum_i \bm{p}_i = N (\bm{v}_G + K \bm{m})$, and the total angular momentum 
 $\bm{L} = \sum_i \left[\omega_i \hat{\bm{e}}_z + \bm{r}_i\times \bm{v}_i + K \bm{r}_i \times \bm{s}_i\right]$, where $\bm{v}_G = N^{-1} \sum_i \dot{\bm{r}}_i$ is the center of mass velocity, $\bm{m} =N^{-1} \sum_i \bm{s}_i$ is the intensive magnetization, and $\hat{\bm{e}}_z = \hat{\bm{e}}_x \times \hat{\bm{e}}_y$ is the out-of-plane unit vector. We perform Molecular Dynamics (MD) simulations in the $(N,V,E)$ ensemble, using initial conditions such that $\bm{P} = \bm{L}= \bm{0}$. 
For  $K = 0$, and at sufficiently low packing fraction, $\phi$, and temperature, $T$,  the system undergoes a ferromagnetism-induced phase separation (FIPS) between a ferromagnetic liquid and a paramagnetic gas due to the emergence of an effective attraction generated by the tendency of the spins to align~\cite{Casiulis2019}.
For $K>0$, a similar phase separation is observed~\cite{Casiulis2019b} (see also the SI). The conservation of momentum imposes that magnetized phases move collectively with velocity $\bm{v}_G = - K \bm{m}$. As a result, for large values of $N$ or $K$, the high kinetic energy cost of the polar moving states prohibits the existence of magnetized phases, and still patterns, such as vortices or solitons, with locally aligned spins but no global magnetization, emerge~\cite{Casiulis2019b}. On the contrary, for small enough $K$ and $N$, the system maintains its magnetization, and spontaneously develops a mean velocity for a wide range of $\phi$ and $T$, see Fig.~\ref{fig:HamFlock}(a). In the following we focus on this last situation.

In the coexistence regime, the moving phase consists in a moving droplet, surrounded by a still gas. Figure~\ref{fig:HamFlock}(b) displays a typical snapshot of the displacements $\bm{u}_i \equiv \bm{r}_i(\tau) - \bm{r}_i(0)$, where $\tau = 2\cdot 10^2$, is chosen long enough to average out thermal fluctuations and reach a  limit in which the correlations are $\tau$ independent (see SI). 
One can observe the strong polarization of the displacements inside the droplet; their polarity, as defined in the context of bird flocks~\cite{Cavagna2010,Hemelrijk2014} is $\Phi \equiv N^{-1} \left| \sum_i \bm{u}_i / \left| \bm{u}_i \right| \right| \simeq 0.95$.
More intriguing are the displacement fluctuations, $\bm{u}_i^\star \equiv \bm{u}_i - \bm{u}_G$, where $\bm{u}_G = N^{-1} \sum_k \bm{u}_k$, displayed in Fig.~\ref{fig:HamFlock}(c). They present a clear spatial organization, which we quantify by computing their spatial correlations across the flock:
\begin{equation}
C_d (r) = \frac{1}{c_0} \frac{\sum\limits_{j\neq i} \bm{u}^\star_i \cdot \bm{u}^\star_j \delta(r - r_{ij}) }{ \sum\limits_{j\neq i} \delta(r - r_{ij})} \; .
\end{equation}
with $\delta$ a binning function~\footnote{See the SI for effect of $\tau$ on the shape of the correlations.} and $c_0$ fixing $C_d(0)=1$.
Similarly, one defines the speed-speed correlation, $C_{sp}(r)$, 
by replacing $\bm{u}_i^\star$ by $\left|\bm{u}_i\right| - N^{-1} \sum \left|\bm{u}_k\right| $ in the above definition.
Figure~\ref{fig:HamFlock}(d)-(e) display $C_d(r)$ and $C_{sp}(r)$ against $r/r_{max}$ for $N=128, 512, 2048$ and $K=0.03, 0.03, 0.001$, respectively, with $r_{max}$ the largest distance between two particles in the droplet.
We observe an excellent data collapse, indicating that the only relevant scale is the droplet size.

Such correlations, without any scale other than the system size, were first reported in bird flocks~\cite{Cavagna2010}. In the theoretical framework of the Vicsek model~\cite{Vicsek1995} and its hydrodynamic description by Toner-Tu~\cite{Toner95,Tu1998,Toner2005}, in which flocking results from the build-up of a true long-range polar order, these correlations have been described as being scale-free, suggesting an underlying critical phenomenon.
Yet, as stated in Ref.~\cite{Cavagna2010}, while the hydrodynamic theories of flocking are good candidates to explain the correlations of the displacement \textit{vectors}, in particular the ones of their orientations, they do not explain the scale-free nature of the \textit{speed} correlations. Explaining the latter within a critical framework requires either the application of an external dynamical field~\cite{Cavagna2013a}, or the introduction of a free energy with a marginal direction~\cite{Cavagna2019}. 

Although it is clear that the present model is \textit{not} a model of birds, with metric instead of topological interactions which 
induce structural features absent in flocks~\cite{Cavagna2008,Ballerini2008a}, the fact that we recover strikingly similar correlations suggests the possible 
contribution of yet another source of correlations, which we shall now unveil.
\begin{figure*}
\centering
\includegraphics[width = 0.9\textwidth]{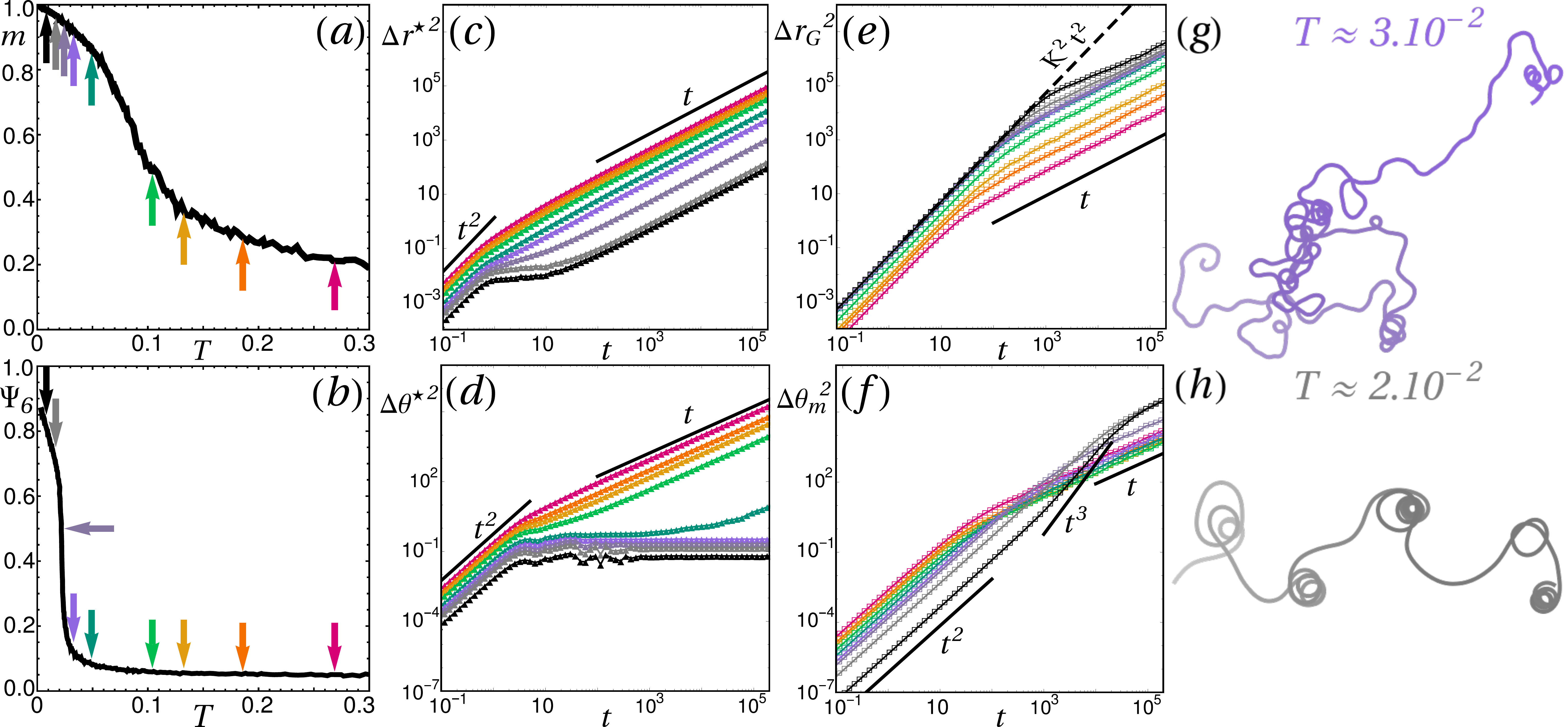}
\vspace{-2mm}
\caption{ {\textbf{Dynamics and Structure of the Homogeneous phases across the Curie line and hexatic transition.}} 
$(a)$ Mean magnetization modulus and $(b)$ hexatic order parameter versus $T$. 
The arrows indicate the $T$ values at which the MSD curves in $(c)-(f)$ are recorded.
$(c)$ MSD and $(d)$ MSAD of the relative displacements versus time, in log-log scale. 
$(e)$ MSD of the center of mass (the dashed black line indicates $v_G  = K$), 
and $(f)$ MSAD of the  magnetization vector versus time, in log-log scale. 
$(g)-(h)$ Trajectories of the center of mass, with time going from light to dark, for two temperatures corresponding to the $(g)$ mauve and $(h)$ light gray curves in $(c)-(f)$.
$N = 128, K = 0.23$, $\phi = 0.55$ (red arrow in Fig.~\ref{fig:HamFlock}$(a)$).
}
\label{fig:MSD}
\vspace{-3mm}
\end{figure*}
We start with a quantitative description of the structure and dynamics of the homogeneous moving phase, and then we focus on the inhomogeneous one. To address the first case, we choose $\phi = 0.55$ (orange arrow in Fig.~\ref{fig:HamFlock}(e)) with $N = 128$ and $K = 0.23$. When decreasing the temperature, the system first magnetizes (Fig.~\ref{fig:MSD}(a)) at a crossover taking place at $T\approx10^{-1}$.
At even lower $T$ ($T\leq 2\cdot 10^{-2}$) a structural organization takes place, as testified by the growth of the hexatic order parameter $\Psi_6$~\cite{Halperin1978}, symptomatic of a 2d system approaching a solid phase (Fig.~\ref{fig:MSD}(b)).
The dynamics are then characterized for a few values of $T$, indicated by arrows in Fig.~\ref{fig:MSD}(a)-(b). 
The relative Mean Square Displacements (MSD), ${\Delta r^\star}^2 (t)$, shown in Fig.~\ref{fig:MSD}(c) are the ones of a usual fluid: they cross over from a short-time ballistic regime to a long-time diffusive one, and feature finite plateaus at low $T$s, revealing the freezing of the dynamics as the system becomes solid~\cite{VanMegen1998,Sanchez-Miranda2015}.
The Mean Square Angular Displacements (MSAD) associated to the spin fluctuations around the mean magnetization, ${\Delta \theta^\star}^2 (t)$ shown in Fig.~\ref{fig:MSD}(d), behave like the MSD, except that the plateaus develop at the magnetization crossover: spin-wave excitations are exponentially suppressed as the temperature decreases~\cite{Tobochnik1979}, similarly to what we found for $K=0$~\cite{Casiulis2019}.
The MSD associated to $\bm{u}_G$, $\Delta r_G^2 (t)$, or ``center-of-mass'' MSD, characterizes the collective motion of the fluid (Fig.~\ref{fig:MSD}(e)). At short times the motion is ballistic, with a collective speed $v_G = K m$ due to momentum conservation. 
At long times it becomes diffusive, with a diffusion coefficient that grows as $T$ decreases. At very low $T$, one notices the presence of a short sub-diffusive regime, which separates the ballistic from the diffusive regime.
Finally, the MSAD $\Delta \theta_m^2 (t)$ associated to the orientation of the magnetization, $\theta_m = \text{atan}(m_y / m_x)$ (Fig.~\ref{fig:MSD}(f)) is ballistic at short times, showing that the magnetization follows inertial dynamics, and diffusive at long times, with a diffusion constant which decreases with decreasing $T$ as long as the system is paramagnetic, and increases when further decreasing $T$ into the ferromagnetic phase.
At very low $T$, and unusual intermediate super-ballistic regime develops at the liquid-solid crossover, 
concomitantly with the sub-diffusive regime of $\Delta r_G^2$. 
The trajectories of the center of mass give a first hint
of its origin. In the isotropic liquid, Fig.~\ref{fig:MSD}(g), the trajectories look like persistent random walks whereas, as the system rigidifies, Fig.~\ref{fig:MSD}(h), loops with accelerated rotations of $\bm{m}$  
appear, leading to a less efficient exploration of space.

The role played by the onset of rigidity can be rationalized, writing down an effective model for the dynamics of the center of mass. We recall that the angular momentum is conserved. Decomposing the positions, spins and velocities as $\bm{r}_i = \bm{r}_G + \bm{r}_i^\star$, $\bm{s}_i = \bm{m} + \bm{\delta s}_i$ and $\bm{v}_i = - K \bm{m} + r_i^\star \Omega \hat{\bm{e}}_{\phi,i} + \bm{\delta v}_{i}$, with $\Omega$ the mean rotational velocity and $\hat{\bm{e}}_{\phi,i} = \hat{\bm{e}}_z \times \bm{r}_i/r_i$, and using that $\bm{P} = \bm{0}$, the angular momentum reads
\begin{align}
\bm{L} = N \dot{\theta}_m \hat{\bm{e}}_z + I \Omega \hat{\bm{e}}_z + \bm{\delta L}\;, 
\label{eq:AngMomentum}
\end{align}
where $I = \sum r_i^2$ is the moment of inertia, and $\bm{\delta L} = \sum \bm{r}_i^\star \times (\bm{\delta v}_i + K \bm{\delta s}_i)$.
Assuming that $\Omega \approx f \dot{\theta}_m$, with $f$ a constant, approximating $\bm{\delta \dot{L}}$ by a noise term, and introducing a characteristic time $\tau_m \sim (N + f I)/(f \dot{I})$, that we treat as a constant, the conservation law $\dot{\bm{L}}=\bm{0}$ reads
\begin{align}
    \tau_m \ddot{\theta}_m + \dot{\theta}_m &= \sqrt{D_r}\eta_m
    \; , \label{eq:LangevinG2}
\end{align}
where $D_r$ is a rotational diffusion constant, and $\eta_m$ a centered Gaussian white noise.
All in all, neglecting for simplicity the fluctuations of the magnetization amplitude, the dynamics of the center of mass obey the equation $\bm{v}_G = - K m(T) \hat{\bm{e}}_{\bm{m}}$ where $\hat{\bm{e}}_{\bm{m}}$, the orientation of $\bm{m}$, 
follows the dynamics prescribed by Eq.~(\ref{eq:LangevinG2}).
\begin{figure}[t!]
    \centering
    \includegraphics[height = 0.45\columnwidth]{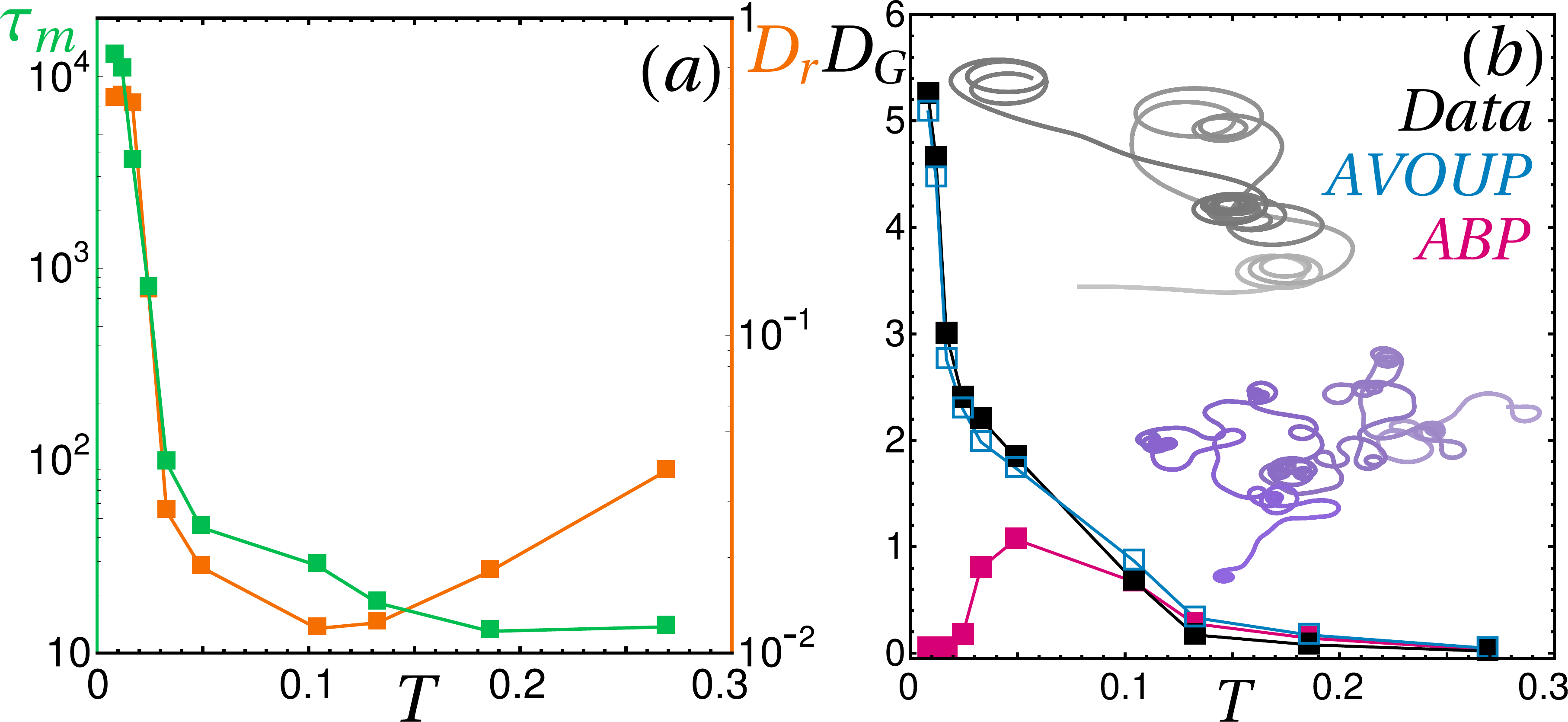}
    \vspace{-1mm}
    \caption{ {\textbf{AVOUP Dynamics.}} 
    $(a)$ $\tau_m$ (green) and $D_r$ (orange) versus  $T$ in lin-log scale.
    $(b)$ $D_G$ against $T$ (black), together with the ABP (red) and AVOUP (blue) estimations.
    Inset: Trajectories, with time going from light to dark, 
    obtained by integrating the AVOUP equations of motion 
    over  $t=10000$, with an integration step $\delta t = 1$, and fixed values $K m = 0.1, D_r = 1$. 
    $\tau_m = 5000$ (gray), and $\tau_m = 500$ (mauve). 
    }
    \label{fig:MemTraj}
    \vspace{-5mm}
\end{figure}
This equation states that $\theta_m$ is subject to inertia and rotational diffusion and defines an Angular Velocity Ornstein-Uhlenbeck Particle (AVOUP) process, a model similar to the ones used to describe experimental living systems, 
ranging from microtubules to flatworms~\cite{Sumino2012,Nagai2015,Chen2017,Sugi2019}. Here, the inertial time $\tau_m$ not only grows as the moment of inertia increases, but also as the deformation dynamics (encoded by $\dot{I}$) are suppressed: in other words, it increases as the system becomes rigid. The associated accelerated dynamics of $\theta_m$ is responsible for the super-ballistic regime observed for $\Delta \theta_m(t)$ and, in turn, the sub-diffusive regime for $\Delta r_G(t)$.
In fact, the large inertia limit, $\tau_m \to \infty$, yields a random-acceleration process, 
that behaves like $\Delta\theta_m^2 \sim t^3$ at long times~\cite{Burkhardt2007}. This regime was 
related to looping trajectories in 
a study of self-propulsion with memory~\cite{Kranz2019}.
We further validate our effective description by comparing the diffusion constant of the center of mass, $D_G$, obtained from the MD simulations, with the theoretical prescription for AVOUP~\cite{Ghosh2015}:
\begin{align}
D_G^{\rm AVOUP}&= D_G^{\rm ABP} \Gamma(\tau_m D_r) \sum\limits_{n = 0}^{\infty} \frac{(\tau_m D_r)^{n+1}}{\Gamma(n + 1 + \tau_m D_r)}
\; ,\label{eq:EffDiff}
\end{align}
where we introduced the $\Gamma$-function and $D_G^{\rm ABP} = K^2 m^2 /(2 D_r)$ 
is the diffusion constant in the over-damped limit of Active Brownian Particles (ABP)~\cite{Fily2012,Winkler2015}. We extract $D_r$ from the long-time diffusive behavior of $\theta_m$, and $\tau_m$ from a short-time exponential fit of the two-time autocorrelation $C(t) = \langle \dot{\theta}_m(0) \dot{\theta}_m(t)\rangle$.
These inputs are plotted against $T$ in Fig.~\ref{fig:MemTraj}(a). Note that $\tau_m$ rises concomitantly with $\Psi_6$, confirming that the inertia of $\theta_m$ is tightly related to the rigidity of the system.
Figure~\ref{fig:MemTraj}(b) shows how good the AVOUP estimation of $D_G$ is, while the zero-memory limit is essentially valid 
only at high $T$, where the magnetization is low. The trajectories obtained by integrating the AVOUP dynamics 
are also good reproductions of the ones shown in Fig.~\ref{fig:MSD}.
%
\begin{figure}[h]
    \centering
    \includegraphics[width = .98\columnwidth]{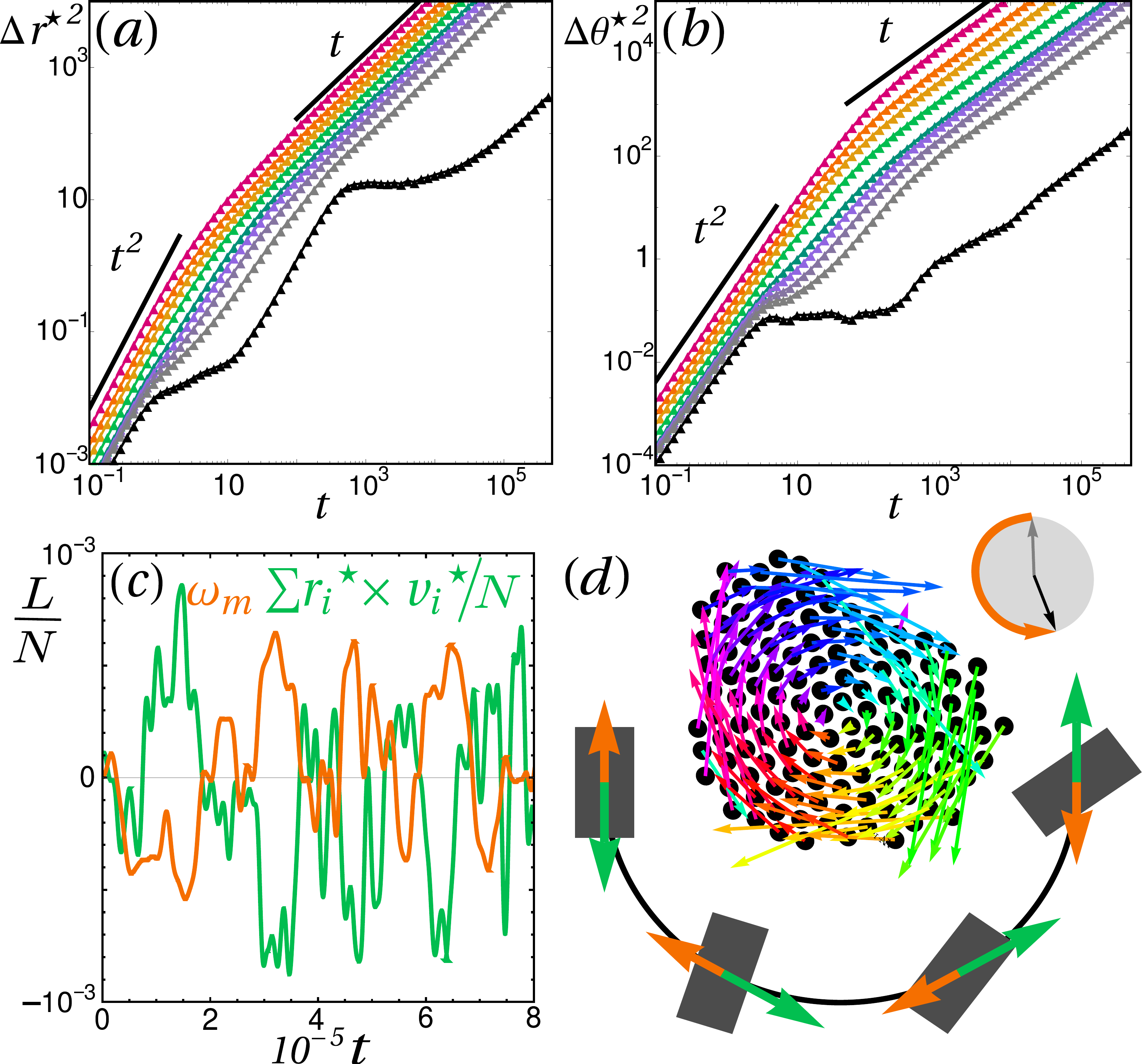}
    \vspace{-2mm}
    \caption{ {\textbf{Spontaneous Rigid Rotations.}}
    Relative $(a)$ MSD and $(b)$ MSAD for $N = 128, K = 0.03, \, \phi \approx 0.14$ (blue arrow in Fig.~\ref{fig:HamFlock}$(a)$) and $T$ increasing from black to red.
    The coexistence line is crossed between the green and teal lines.
    $(c)$ Typical $\omega_m = \dot{\theta}_m$ (orange) and $N^{-1} \sum\bm{r}_i^\star \times \bm{v}_i^\star$ (green), time-averaged over 
    $\tau_w = 10^2$, versus time ($T \approx 10^{-2}$).
    $(d)$ Relative displacements color-coded according to their orientation and rescaled by their mean ($T \approx 10^{-2}, \tau \approx 10^4$), together with a sketch of the spontaneous rotation dynamics ($\bm{m}$ in orange and $\bm{v}_G$ in green).  Upper inset: the rotation of $\bm{m}$ from the gray  to the black arrow, following the orange curved arrow.
}
    \label{fig:AngMom}
    \vspace{-6mm}
\end{figure}

Coming back to the phase-separated states, say at $\phi = 0.14$, $N = 128$, and $K =$ $0.03$, we observe that
the AVOUP description and the general features of the center of mass trajectories are almost unchanged, see Fig.~\ref{fig:AngMom}.
The only major difference occurs at very low $T$, where the droplet is highly magnetized, structurally ordered, dynamically frozen, and immersed in a very low density gas.
Under these circumstances, the relative displacements exhibit an intermediate second ballistic regime, that can also be explained by the conservation of ${\bm{L}}$.
As an individual flock is free to rotate around its center of mass, a rotation of the magnetization must be compensated by either a rigid rotation in the opposite direction or thermal fluctuations [see Eq.~(\ref{eq:AngMomentum})].
At very low $T$, the thermal fluctuations are no longer enough to absorb changes in $\dot{\theta}_m$. As a result, for $\bm{L} = \bm{0}$, the droplets develop rigid rotations such that $N \dot{\theta}_m = - I \Omega$, as confirmed when plotting time-averaged versions of $\dot{\theta}_m$ and $N^{-1} \sum\bm{r}_i^\star \times \bm{v}_i^\star$, Fig.~\ref{fig:AngMom}(c).
The associated displacement fields, Fig.~\ref{fig:AngMom}(d), explain our initial observations, Fig.~\ref{fig:HamFlock}(d)-(e): a rigid rotation yields a correlation of the relative displacements at the scale of the size of the system (see also \textit{Movie2.mp4} in the SI). The excellent match between 
the numerically observed and 
analytical $C_d$ and $C_{sp}$ for a homogeneous rigid disk rotating at a constant rate (see the SI), 
shown with black lines in Fig.~\ref{fig:HamFlock}(d)-(e), confirms the above interpretation.

We showed that Hamiltonian flocks exhibit correlations remarkably similar to those observed in bird flocks, with no other scale than the droplet size.
In the latter, the scaling and functional form of the correlations stem from rigid body rotations around the center of mass of the flock.
Hence the question: in bird flocks, what is the part of the correlations due to,
on the one hand, the Goldstone mode along the orthoradial direction of an effective potential ($C_d$)~\cite{Cavagna2010} and the marginality of its radial direction ($C_{sp}$)~\cite{Cavagna2019}
and, on the other hand, rotations? 
It is not obvious which effect is stronger. 
For physiological reasons, the rotations cannot be large~\cite{Cavagna2018} but in an $O(n)$ model, the amplitude of the correlation scales like $1-m^2$, with $m$ the magnetization, and is thus also very small for well polarized systems like bird flocks~\cite{Cavagna2010}.
To assess the importance of each effect, we analyze the data from the SI of Ref.~\cite{Cavagna2010} (see SI): while the shape of $C_d$ is dominated by the spin wave contribution,
$C_{sp}$ is controlled by the rigid body rotation. This deserves a few comments.
First, a rotation of a fraction of a degree, compatible with the variability of bird speeds~\cite{Cavagna2010}, is sufficient to set the shape of $C_{sp}$.
Second, although flocks do not turn via a rotation around an external point~\cite{Cavagna2018}, a small rotation around the center of mass is not precluded: in fact, a turn due to a spin wave generically contains a small, but finite rotation.
Thus, while rigid rotations are \textit{not} entirely responsible for velocity correlations in bird flocks, they may take part in them, and should not be overlooked.
In particular, the variability in the shape of $C_{sp}$ across bird flocks~\cite{CavagnaComment} calls for a more systematic analysis.

Finally, we stress that, in Hamiltonian flocks, the rotation is rooted in the angular momentum conservation. Yet, rigid body rotations were also reported in systems  with no strict angular momentum conservation, like self-propelled Janus colloids~\cite{Ginot2018,VanderLinden2019} and active dumbbells~\cite{Petrelli2018}.
Whether the rotational invariance of active systems leads to pseudo-conserved generalized angular momenta remains to be elucidated.

\acknowledgements{We thank A. Jeli\'c for useful discussions as well as I. Giardina and A. Cavagna for their careful reading of, and rich comments on, our work.
\vspace{-5mm}
}

\bibliography{Bibtex-Droplets-LC}

\begin{thebibliography}{53}%
\makeatletter
\providecommand \@ifxundefined [1]{%
 \@ifx{#1\undefined}
}%
\providecommand \@ifnum [1]{%
 \ifnum #1\expandafter \@firstoftwo
 \else \expandafter \@secondoftwo
 \fi
}%
\providecommand \@ifx [1]{%
 \ifx #1\expandafter \@firstoftwo
 \else \expandafter \@secondoftwo
 \fi
}%
\providecommand \natexlab [1]{#1}%
\providecommand \enquote  [1]{``#1''}%
\providecommand \bibnamefont  [1]{#1}%
\providecommand \bibfnamefont [1]{#1}%
\providecommand \citenamefont [1]{#1}%
\providecommand \href@noop [0]{\@secondoftwo}%
\providecommand \href [0]{\begingroup \@sanitize@url \@href}%
\providecommand \@href[1]{\@@startlink{#1}\@@href}%
\providecommand \@@href[1]{\endgroup#1\@@endlink}%
\providecommand \@sanitize@url [0]{\catcode `\\12\catcode `\$12\catcode
  `\&12\catcode `\#12\catcode `\^12\catcode `\_12\catcode `\%12\relax}%
\providecommand \@@startlink[1]{}%
\providecommand \@@endlink[0]{}%
\providecommand \url  [0]{\begingroup\@sanitize@url \@url }%
\providecommand \@url [1]{\endgroup\@href {#1}{\urlprefix }}%
\providecommand \urlprefix  [0]{URL }%
\providecommand \Eprint [0]{\href }%
\providecommand \doibase [0]{http://dx.doi.org/}%
\providecommand \selectlanguage [0]{\@gobble}%
\providecommand \bibinfo  [0]{\@secondoftwo}%
\providecommand \bibfield  [0]{\@secondoftwo}%
\providecommand \translation [1]{[#1]}%
\providecommand \BibitemOpen [0]{}%
\providecommand \bibitemStop [0]{}%
\providecommand \bibitemNoStop [0]{.\EOS\space}%
\providecommand \EOS [0]{\spacefactor3000\relax}%
\providecommand \BibitemShut  [1]{\csname bibitem#1\endcsname}%
\let\auto@bib@innerbib\@empty
\bibitem [{\citenamefont {{Breder Jr.}}(1954)}]{Breder1954}%
  \BibitemOpen
  \bibfield  {author} {\bibinfo {author} {\bibfnamefont {C.~M.}\ \bibnamefont
  {{Breder Jr.}}},\ }\href@noop {} {\bibfield  {journal} {\bibinfo  {journal}
  {Ecology}\ }\textbf {\bibinfo {volume} {35}},\ \bibinfo {pages} {361}
  (\bibinfo {year} {1954})}\BibitemShut {NoStop}%
\bibitem [{\citenamefont {Aoki}(1982)}]{Aoki1982}%
  \BibitemOpen
  \bibfield  {author} {\bibinfo {author} {\bibfnamefont {I.}~\bibnamefont
  {Aoki}},\ }\href@noop {} {\bibfield  {journal} {\bibinfo  {journal} {Nippon
  Suisan Gakkaishi}\ }\textbf {\bibinfo {volume} {48}},\ \bibinfo {pages}
  {1081} (\bibinfo {year} {1982})}\BibitemShut {NoStop}%
\bibitem [{\citenamefont {Badgerow}(1988)}]{Badgerow1988}%
  \BibitemOpen
  \bibfield  {author} {\bibinfo {author} {\bibfnamefont {J.~P.}\ \bibnamefont
  {Badgerow}},\ }\href@noop {} {\bibfield  {journal} {\bibinfo  {journal}
  {Auk}\ }\textbf {\bibinfo {volume} {105}},\ \bibinfo {pages} {749} (\bibinfo
  {year} {1988})}\BibitemShut {NoStop}%
\bibitem [{\citenamefont {Huth}\ and\ \citenamefont {Wissel}(1992)}]{Huth1992}%
  \BibitemOpen
  \bibfield  {author} {\bibinfo {author} {\bibfnamefont {A.}~\bibnamefont
  {Huth}}\ and\ \bibinfo {author} {\bibfnamefont {C.}~\bibnamefont {Wissel}},\
  }\href@noop {} {\bibfield  {journal} {\bibinfo  {journal} {J. Theor. Biol.}\
  }\textbf {\bibinfo {volume} {156}},\ \bibinfo {pages} {365} (\bibinfo {year}
  {1992})}\BibitemShut {NoStop}%
\bibitem [{\citenamefont {Krause}\ and\ \citenamefont
  {Ruxton}(2002)}]{Krause2002}%
  \BibitemOpen
  \bibfield  {author} {\bibinfo {author} {\bibfnamefont {J.}~\bibnamefont
  {Krause}}\ and\ \bibinfo {author} {\bibfnamefont {G.~D.}\ \bibnamefont
  {Ruxton}},\ }\href@noop {} {\emph {\bibinfo {title} {{Living in Groups}}}}\
  (\bibinfo  {publisher} {Oxford University Press},\ \bibinfo {address}
  {Oxford},\ \bibinfo {year} {2002})\BibitemShut {NoStop}%
\bibitem [{\citenamefont {Tunstr{\o}m}\ \emph {et~al.}(2013)\citenamefont
  {Tunstr{\o}m}, \citenamefont {Katz}, \citenamefont {Ioannou}, \citenamefont
  {Huepe}, \citenamefont {Lutz},\ and\ \citenamefont {Couzin}}]{Tunstrom2013}%
  \BibitemOpen
  \bibfield  {author} {\bibinfo {author} {\bibfnamefont {K.}~\bibnamefont
  {Tunstr{\o}m}}, \bibinfo {author} {\bibfnamefont {Y.}~\bibnamefont {Katz}},
  \bibinfo {author} {\bibfnamefont {C.~C.}\ \bibnamefont {Ioannou}}, \bibinfo
  {author} {\bibfnamefont {C.}~\bibnamefont {Huepe}}, \bibinfo {author}
  {\bibfnamefont {M.~J.}\ \bibnamefont {Lutz}}, \ and\ \bibinfo {author}
  {\bibfnamefont {I.~D.}\ \bibnamefont {Couzin}},\ }\href@noop {} {\bibfield
  {journal} {\bibinfo  {journal} {PLoS Comput. Biol.}\ }\textbf {\bibinfo
  {volume} {9}},\ \bibinfo {pages} {e1002915} (\bibinfo {year}
  {2013})}\BibitemShut {NoStop}%
\bibitem [{\citenamefont {Rosenthal}\ \emph {et~al.}(2015)\citenamefont
  {Rosenthal}, \citenamefont {Twomey}, \citenamefont {Hartnett}, \citenamefont
  {Wu},\ and\ \citenamefont {Couzin}}]{Rosenthal2015}%
  \BibitemOpen
  \bibfield  {author} {\bibinfo {author} {\bibfnamefont {S.~B.}\ \bibnamefont
  {Rosenthal}}, \bibinfo {author} {\bibfnamefont {C.~R.}\ \bibnamefont
  {Twomey}}, \bibinfo {author} {\bibfnamefont {A.~T.}\ \bibnamefont
  {Hartnett}}, \bibinfo {author} {\bibfnamefont {H.~S.}\ \bibnamefont {Wu}}, \
  and\ \bibinfo {author} {\bibfnamefont {I.~D.}\ \bibnamefont {Couzin}},\
  }\href@noop {} {\bibfield  {journal} {\bibinfo  {journal} {Proc. Natl. Acad.
  Sci.}\ }\textbf {\bibinfo {volume} {112}},\ \bibinfo {pages} {4690} (\bibinfo
  {year} {2015})}\BibitemShut {NoStop}%
\bibitem [{\citenamefont {Vicsek}\ \emph {et~al.}(1995)\citenamefont {Vicsek},
  \citenamefont {Czir{\'{o}}k}, \citenamefont {Ben-Jacob}, \citenamefont
  {Cohen},\ and\ \citenamefont {Shochet}}]{Vicsek1995}%
  \BibitemOpen
  \bibfield  {author} {\bibinfo {author} {\bibfnamefont {T.}~\bibnamefont
  {Vicsek}}, \bibinfo {author} {\bibfnamefont {A.}~\bibnamefont
  {Czir{\'{o}}k}}, \bibinfo {author} {\bibfnamefont {E.}~\bibnamefont
  {Ben-Jacob}}, \bibinfo {author} {\bibfnamefont {I.}~\bibnamefont {Cohen}}, \
  and\ \bibinfo {author} {\bibfnamefont {O.}~\bibnamefont {Shochet}},\
  }\href@noop {} {\bibfield  {journal} {\bibinfo  {journal} {Phys. Rev. Lett.}\
  }\textbf {\bibinfo {volume} {75}},\ \bibinfo {pages} {1226} (\bibinfo {year}
  {1995})}\BibitemShut {NoStop}%
\bibitem [{\citenamefont {Gr{\'{e}}goire}\ and\ \citenamefont
  {Chat{\'{e}}}(2004)}]{Gregoire2004}%
  \BibitemOpen
  \bibfield  {author} {\bibinfo {author} {\bibfnamefont {G.}~\bibnamefont
  {Gr{\'{e}}goire}}\ and\ \bibinfo {author} {\bibfnamefont {H.}~\bibnamefont
  {Chat{\'{e}}}},\ }\href@noop {} {\bibfield  {journal} {\bibinfo  {journal}
  {Phys. Rev. Lett.}\ }\textbf {\bibinfo {volume} {92}},\ \bibinfo {pages}
  {025702} (\bibinfo {year} {2004})}\BibitemShut {NoStop}%
\bibitem [{\citenamefont {Vicsek}\ and\ \citenamefont
  {Zafeiris}(2012)}]{Vicsek2012}%
  \BibitemOpen
  \bibfield  {author} {\bibinfo {author} {\bibfnamefont {T.}~\bibnamefont
  {Vicsek}}\ and\ \bibinfo {author} {\bibfnamefont {A.}~\bibnamefont
  {Zafeiris}},\ }\href@noop {} {\bibfield  {journal} {\bibinfo  {journal}
  {Phys. Rep.}\ }\textbf {\bibinfo {volume} {517}},\ \bibinfo {pages} {71}
  (\bibinfo {year} {2012})}\BibitemShut {NoStop}%
\bibitem [{\citenamefont {Toner}\ and\ \citenamefont {Tu}(1995)}]{Toner95}%
  \BibitemOpen
  \bibfield  {author} {\bibinfo {author} {\bibfnamefont {J.}~\bibnamefont
  {Toner}}\ and\ \bibinfo {author} {\bibfnamefont {Y.}~\bibnamefont {Tu}},\
  }\href@noop {} {\bibfield  {journal} {\bibinfo  {journal} {Phys. Rev. Lett.}\
  }\textbf {\bibinfo {volume} {75}},\ \bibinfo {pages} {4326} (\bibinfo {year}
  {1995})}\BibitemShut {NoStop}%
\bibitem [{\citenamefont {Tu}\ \emph {et~al.}(1998)\citenamefont {Tu},
  \citenamefont {Toner},\ and\ \citenamefont {Ulm}}]{Tu1998}%
  \BibitemOpen
  \bibfield  {author} {\bibinfo {author} {\bibfnamefont {Y.}~\bibnamefont
  {Tu}}, \bibinfo {author} {\bibfnamefont {J.}~\bibnamefont {Toner}}, \ and\
  \bibinfo {author} {\bibfnamefont {M.}~\bibnamefont {Ulm}},\ }\href@noop {}
  {\bibfield  {journal} {\bibinfo  {journal} {Phys. Rev. Lett.}\ }\textbf
  {\bibinfo {volume} {80}},\ \bibinfo {pages} {4819} (\bibinfo {year}
  {1998})}\BibitemShut {NoStop}%
\bibitem [{\citenamefont {Toner}\ \emph {et~al.}(2005)\citenamefont {Toner},
  \citenamefont {Tu},\ and\ \citenamefont {Ramaswamy}}]{Toner2005}%
  \BibitemOpen
  \bibfield  {author} {\bibinfo {author} {\bibfnamefont {J.}~\bibnamefont
  {Toner}}, \bibinfo {author} {\bibfnamefont {Y.}~\bibnamefont {Tu}}, \ and\
  \bibinfo {author} {\bibfnamefont {S.}~\bibnamefont {Ramaswamy}},\ }\href@noop
  {} {\bibfield  {journal} {\bibinfo  {journal} {Ann. Phys. (N. Y).}\ }\textbf
  {\bibinfo {volume} {318}},\ \bibinfo {pages} {170} (\bibinfo {year}
  {2005})}\BibitemShut {NoStop}%
\bibitem [{\citenamefont {Peshkov}\ \emph {et~al.}(2014)\citenamefont
  {Peshkov}, \citenamefont {Bertin}, \citenamefont {Ginelli},\ and\
  \citenamefont {Chat{\'{e}}}}]{Peshkov2014}%
  \BibitemOpen
  \bibfield  {author} {\bibinfo {author} {\bibfnamefont {A.}~\bibnamefont
  {Peshkov}}, \bibinfo {author} {\bibfnamefont {E.}~\bibnamefont {Bertin}},
  \bibinfo {author} {\bibfnamefont {F.}~\bibnamefont {Ginelli}}, \ and\
  \bibinfo {author} {\bibfnamefont {H.}~\bibnamefont {Chat{\'{e}}}},\
  }\href@noop {} {\bibfield  {journal} {\bibinfo  {journal} {Eur. Phys. J.
  Spec. Top.}\ }\textbf {\bibinfo {volume} {223}},\ \bibinfo {pages} {1315}
  (\bibinfo {year} {2014})}\BibitemShut {NoStop}%
\bibitem [{\citenamefont {Marchetti}\ \emph {et~al.}(2013)\citenamefont
  {Marchetti}, \citenamefont {Joanny}, \citenamefont {Ramaswamy}, \citenamefont
  {Liverpool}, \citenamefont {Prost}, \citenamefont {Rao},\ and\ \citenamefont
  {Simha}}]{Marchetti2013}%
  \BibitemOpen
  \bibfield  {author} {\bibinfo {author} {\bibfnamefont {M.~C.}\ \bibnamefont
  {Marchetti}}, \bibinfo {author} {\bibfnamefont {J.~F.}\ \bibnamefont
  {Joanny}}, \bibinfo {author} {\bibfnamefont {S.}~\bibnamefont {Ramaswamy}},
  \bibinfo {author} {\bibfnamefont {T.~B.}\ \bibnamefont {Liverpool}}, \bibinfo
  {author} {\bibfnamefont {J.}~\bibnamefont {Prost}}, \bibinfo {author}
  {\bibfnamefont {M.}~\bibnamefont {Rao}}, \ and\ \bibinfo {author}
  {\bibfnamefont {R.~A.}\ \bibnamefont {Simha}},\ }\href@noop {} {\bibfield
  {journal} {\bibinfo  {journal} {Rev. Mod. Phys.}\ }\textbf {\bibinfo {volume}
  {85}},\ \bibinfo {pages} {1143(47)} (\bibinfo {year} {2013})}\BibitemShut
  {NoStop}%
\bibitem [{\citenamefont {Yang}\ and\ \citenamefont
  {Marchetti}(2015)}]{Yang2015}%
  \BibitemOpen
  \bibfield  {author} {\bibinfo {author} {\bibfnamefont {X.}~\bibnamefont
  {Yang}}\ and\ \bibinfo {author} {\bibfnamefont {M.~C.}\ \bibnamefont
  {Marchetti}},\ }\href@noop {} {\bibfield  {journal} {\bibinfo  {journal}
  {Phys. Rev. Lett.}\ }\textbf {\bibinfo {volume} {115}},\ \bibinfo {pages}
  {258101} (\bibinfo {year} {2015})}\BibitemShut {NoStop}%
\bibitem [{\citenamefont {Schaller}\ \emph {et~al.}(2010)\citenamefont
  {Schaller}, \citenamefont {Weber}, \citenamefont {Semmrich}, \citenamefont
  {Frey},\ and\ \citenamefont {Bausch}}]{Schaller2010}%
  \BibitemOpen
  \bibfield  {author} {\bibinfo {author} {\bibfnamefont {V.}~\bibnamefont
  {Schaller}}, \bibinfo {author} {\bibfnamefont {C.}~\bibnamefont {Weber}},
  \bibinfo {author} {\bibfnamefont {C.}~\bibnamefont {Semmrich}}, \bibinfo
  {author} {\bibfnamefont {E.}~\bibnamefont {Frey}}, \ and\ \bibinfo {author}
  {\bibfnamefont {A.~R.}\ \bibnamefont {Bausch}},\ }\href@noop {} {\bibfield
  {journal} {\bibinfo  {journal} {Nature}\ }\textbf {\bibinfo {volume} {467}},\
  \bibinfo {pages} {73} (\bibinfo {year} {2010})}\BibitemShut {NoStop}%
\bibitem [{\citenamefont {Deseigne}\ \emph {et~al.}(2010)\citenamefont
  {Deseigne}, \citenamefont {Dauchot},\ and\ \citenamefont
  {Chat{\'{e}}}}]{Deseigne2010}%
  \BibitemOpen
  \bibfield  {author} {\bibinfo {author} {\bibfnamefont {J.}~\bibnamefont
  {Deseigne}}, \bibinfo {author} {\bibfnamefont {O.}~\bibnamefont {Dauchot}}, \
  and\ \bibinfo {author} {\bibfnamefont {H.}~\bibnamefont {Chat{\'{e}}}},\
  }\href@noop {} {\bibfield  {journal} {\bibinfo  {journal} {Phys. Rev. Lett.}\
  }\textbf {\bibinfo {volume} {105}},\ \bibinfo {pages} {098001} (\bibinfo
  {year} {2010})}\BibitemShut {NoStop}%
\bibitem [{\citenamefont {Bricard}\ \emph {et~al.}(2013)\citenamefont
  {Bricard}, \citenamefont {Caussin}, \citenamefont {Desreumaux}, \citenamefont
  {Dauchot},\ and\ \citenamefont {Bartolo}}]{Bricard2013}%
  \BibitemOpen
  \bibfield  {author} {\bibinfo {author} {\bibfnamefont {A.}~\bibnamefont
  {Bricard}}, \bibinfo {author} {\bibfnamefont {J.-B.}\ \bibnamefont
  {Caussin}}, \bibinfo {author} {\bibfnamefont {N.}~\bibnamefont {Desreumaux}},
  \bibinfo {author} {\bibfnamefont {O.}~\bibnamefont {Dauchot}}, \ and\
  \bibinfo {author} {\bibfnamefont {D.}~\bibnamefont {Bartolo}},\ }\href@noop
  {} {\bibfield  {journal} {\bibinfo  {journal} {Nature}\ }\textbf {\bibinfo
  {volume} {503}},\ \bibinfo {pages} {95} (\bibinfo {year} {2013})}\BibitemShut
  {NoStop}%
\bibitem [{\citenamefont {Geyer}\ \emph {et~al.}(2018)\citenamefont {Geyer},
  \citenamefont {Morin},\ and\ \citenamefont {Bartolo}}]{Geyer2018}%
  \BibitemOpen
  \bibfield  {author} {\bibinfo {author} {\bibfnamefont {D.}~\bibnamefont
  {Geyer}}, \bibinfo {author} {\bibfnamefont {A.}~\bibnamefont {Morin}}, \ and\
  \bibinfo {author} {\bibfnamefont {D.}~\bibnamefont {Bartolo}},\ }\href@noop
  {} {\bibfield  {journal} {\bibinfo  {journal} {Nat. Mater.}\ }\textbf
  {\bibinfo {volume} {17}},\ \bibinfo {pages} {789} (\bibinfo {year}
  {2018})}\BibitemShut {NoStop}%
\bibitem [{\citenamefont {Cavagna}\ \emph {et~al.}(2010)\citenamefont
  {Cavagna}, \citenamefont {Cimarelli}, \citenamefont {Giardina}, \citenamefont
  {Parisi}, \citenamefont {Santagati}, \citenamefont {Stefanini},\ and\
  \citenamefont {Viale}}]{Cavagna2010}%
  \BibitemOpen
  \bibfield  {author} {\bibinfo {author} {\bibfnamefont {A.}~\bibnamefont
  {Cavagna}}, \bibinfo {author} {\bibfnamefont {A.}~\bibnamefont {Cimarelli}},
  \bibinfo {author} {\bibfnamefont {I.}~\bibnamefont {Giardina}}, \bibinfo
  {author} {\bibfnamefont {G.}~\bibnamefont {Parisi}}, \bibinfo {author}
  {\bibfnamefont {R.}~\bibnamefont {Santagati}}, \bibinfo {author}
  {\bibfnamefont {F.}~\bibnamefont {Stefanini}}, \ and\ \bibinfo {author}
  {\bibfnamefont {M.}~\bibnamefont {Viale}},\ }\href@noop {} {\bibfield
  {journal} {\bibinfo  {journal} {Proc. Natl. Acad. Sci.}\ }\textbf {\bibinfo
  {volume} {107}},\ \bibinfo {pages} {11865} (\bibinfo {year}
  {2010})}\BibitemShut {NoStop}%
\bibitem [{\citenamefont {Cavagna}\ \emph {et~al.}(2013)\citenamefont
  {Cavagna}, \citenamefont {Giardina},\ and\ \citenamefont
  {Ginelli}}]{Cavagna2013a}%
  \BibitemOpen
  \bibfield  {author} {\bibinfo {author} {\bibfnamefont {A.}~\bibnamefont
  {Cavagna}}, \bibinfo {author} {\bibfnamefont {I.}~\bibnamefont {Giardina}}, \
  and\ \bibinfo {author} {\bibfnamefont {F.}~\bibnamefont {Ginelli}},\
  }\href@noop {} {\bibfield  {journal} {\bibinfo  {journal} {Phys. Rev. Lett.}\
  }\textbf {\bibinfo {volume} {110}},\ \bibinfo {pages} {168107} (\bibinfo
  {year} {2013})}\BibitemShut {NoStop}%
\bibitem [{\citenamefont {Attanasi}\ \emph {et~al.}(2014)\citenamefont
  {Attanasi}, \citenamefont {Cavagna}, \citenamefont {{Del Castello}},
  \citenamefont {Giardina}, \citenamefont {Grigera}, \citenamefont {Jelic},
  \citenamefont {Melillo}, \citenamefont {Parisi}, \citenamefont {Pohl},
  \citenamefont {Shen},\ and\ \citenamefont {Viale}}]{Attanasi2014a}%
  \BibitemOpen
  \bibfield  {author} {\bibinfo {author} {\bibfnamefont {A.}~\bibnamefont
  {Attanasi}}, \bibinfo {author} {\bibfnamefont {A.}~\bibnamefont {Cavagna}},
  \bibinfo {author} {\bibfnamefont {L.}~\bibnamefont {{Del Castello}}},
  \bibinfo {author} {\bibfnamefont {I.}~\bibnamefont {Giardina}}, \bibinfo
  {author} {\bibfnamefont {T.~S.}\ \bibnamefont {Grigera}}, \bibinfo {author}
  {\bibfnamefont {A.}~\bibnamefont {Jelic}}, \bibinfo {author} {\bibfnamefont
  {S.}~\bibnamefont {Melillo}}, \bibinfo {author} {\bibfnamefont
  {L.}~\bibnamefont {Parisi}}, \bibinfo {author} {\bibfnamefont
  {O.}~\bibnamefont {Pohl}}, \bibinfo {author} {\bibfnamefont {E.}~\bibnamefont
  {Shen}}, \ and\ \bibinfo {author} {\bibfnamefont {M.}~\bibnamefont {Viale}},\
  }\href@noop {} {\bibfield  {journal} {\bibinfo  {journal} {Nat. Phys.}\
  }\textbf {\bibinfo {volume} {10}},\ \bibinfo {pages} {691} (\bibinfo {year}
  {2014})}\BibitemShut {NoStop}%
\bibitem [{\citenamefont {Hemelrijk}\ and\ \citenamefont
  {Hildenbrandt}(2015)}]{Hemelrijk2014}%
  \BibitemOpen
  \bibfield  {author} {\bibinfo {author} {\bibfnamefont {C.~K.}\ \bibnamefont
  {Hemelrijk}}\ and\ \bibinfo {author} {\bibfnamefont {H.}~\bibnamefont
  {Hildenbrandt}},\ }\href@noop {} {\bibfield  {journal} {\bibinfo  {journal}
  {J. Stat. Phys.}\ }\textbf {\bibinfo {volume} {158}},\ \bibinfo {pages} {563}
  (\bibinfo {year} {2015})}\BibitemShut {NoStop}%
\bibitem [{\citenamefont {Attanasi}\ \emph {et~al.}(2015)\citenamefont
  {Attanasi}, \citenamefont {Cavagna}, \citenamefont {{Del Castello}},
  \citenamefont {Giardina}, \citenamefont {Jelic}, \citenamefont {Melillo},
  \citenamefont {Parisi}, \citenamefont {Pohl}, \citenamefont {Shen},\ and\
  \citenamefont {Viale}}]{Attanasi2015}%
  \BibitemOpen
  \bibfield  {author} {\bibinfo {author} {\bibfnamefont {A.}~\bibnamefont
  {Attanasi}}, \bibinfo {author} {\bibfnamefont {A.}~\bibnamefont {Cavagna}},
  \bibinfo {author} {\bibfnamefont {L.}~\bibnamefont {{Del Castello}}},
  \bibinfo {author} {\bibfnamefont {I.}~\bibnamefont {Giardina}}, \bibinfo
  {author} {\bibfnamefont {A.}~\bibnamefont {Jelic}}, \bibinfo {author}
  {\bibfnamefont {S.}~\bibnamefont {Melillo}}, \bibinfo {author} {\bibfnamefont
  {L.}~\bibnamefont {Parisi}}, \bibinfo {author} {\bibfnamefont
  {O.}~\bibnamefont {Pohl}}, \bibinfo {author} {\bibfnamefont {E.}~\bibnamefont
  {Shen}}, \ and\ \bibinfo {author} {\bibfnamefont {M.}~\bibnamefont {Viale}},\
  }\href@noop {} {\bibfield  {journal} {\bibinfo  {journal} {J. R. Soc.
  Interface}\ }\textbf {\bibinfo {volume} {12}},\ \bibinfo {pages} {20150319}
  (\bibinfo {year} {2015})}\BibitemShut {NoStop}%
\bibitem [{\citenamefont {Cavagna}\ \emph {et~al.}(2018)\citenamefont
  {Cavagna}, \citenamefont {Giardina},\ and\ \citenamefont
  {Grigera}}]{Cavagna2018}%
  \BibitemOpen
  \bibfield  {author} {\bibinfo {author} {\bibfnamefont {A.}~\bibnamefont
  {Cavagna}}, \bibinfo {author} {\bibfnamefont {I.}~\bibnamefont {Giardina}}, \
  and\ \bibinfo {author} {\bibfnamefont {T.~S.}\ \bibnamefont {Grigera}},\
  }\href@noop {} {\bibfield  {journal} {\bibinfo  {journal} {Phys. Rep.}\
  }\textbf {\bibinfo {volume} {728}},\ \bibinfo {pages} {1} (\bibinfo {year}
  {2018})}\BibitemShut {NoStop}%
\bibitem [{\citenamefont {Cavagna}\ \emph
  {et~al.}(2019{\natexlab{a}})\citenamefont {Cavagna}, \citenamefont {Culla},
  \citenamefont {{Di Carlo}}, \citenamefont {Giardina},\ and\ \citenamefont
  {Grigera}}]{Cavagna2019}%
  \BibitemOpen
  \bibfield  {author} {\bibinfo {author} {\bibfnamefont {A.}~\bibnamefont
  {Cavagna}}, \bibinfo {author} {\bibfnamefont {A.}~\bibnamefont {Culla}},
  \bibinfo {author} {\bibfnamefont {L.}~\bibnamefont {{Di Carlo}}}, \bibinfo
  {author} {\bibfnamefont {I.}~\bibnamefont {Giardina}}, \ and\ \bibinfo
  {author} {\bibfnamefont {T.~S.}\ \bibnamefont {Grigera}},\ }\href@noop {}
  {\bibfield  {journal} {\bibinfo  {journal} {Comptes Rendus Phys.}\ }\textbf
  {\bibinfo {volume} {20}},\ \bibinfo {pages} {319} (\bibinfo {year}
  {2019}{\natexlab{a}})}\BibitemShut {NoStop}%
\bibitem [{\citenamefont {Deseigne}\ \emph {et~al.}(2012)\citenamefont
  {Deseigne}, \citenamefont {L{\'{e}}onard}, \citenamefont {Dauchot},\ and\
  \citenamefont {Chat{\'{e}}}}]{Deseigne2012}%
  \BibitemOpen
  \bibfield  {author} {\bibinfo {author} {\bibfnamefont {J.}~\bibnamefont
  {Deseigne}}, \bibinfo {author} {\bibfnamefont {S.}~\bibnamefont
  {L{\'{e}}onard}}, \bibinfo {author} {\bibfnamefont {O.}~\bibnamefont
  {Dauchot}}, \ and\ \bibinfo {author} {\bibfnamefont {H.}~\bibnamefont
  {Chat{\'{e}}}},\ }\href@noop {} {\bibfield  {journal} {\bibinfo  {journal}
  {Soft Matter}\ }\textbf {\bibinfo {volume} {8}},\ \bibinfo {pages} {5629}
  (\bibinfo {year} {2012})}\BibitemShut {NoStop}%
\bibitem [{\citenamefont {Weber}\ \emph {et~al.}(2013)\citenamefont {Weber},
  \citenamefont {Hanke}, \citenamefont {Deseigne}, \citenamefont
  {L{\'{e}}onard}, \citenamefont {Dauchot}, \citenamefont {Frey},\ and\
  \citenamefont {Chat{\'{e}}}}]{Weber2013}%
  \BibitemOpen
  \bibfield  {author} {\bibinfo {author} {\bibfnamefont {C.~A.}\ \bibnamefont
  {Weber}}, \bibinfo {author} {\bibfnamefont {T.}~\bibnamefont {Hanke}},
  \bibinfo {author} {\bibfnamefont {J.}~\bibnamefont {Deseigne}}, \bibinfo
  {author} {\bibfnamefont {S.}~\bibnamefont {L{\'{e}}onard}}, \bibinfo {author}
  {\bibfnamefont {O.}~\bibnamefont {Dauchot}}, \bibinfo {author} {\bibfnamefont
  {E.}~\bibnamefont {Frey}}, \ and\ \bibinfo {author} {\bibfnamefont
  {H.}~\bibnamefont {Chat{\'{e}}}},\ }\href@noop {} {\bibfield  {journal}
  {\bibinfo  {journal} {Phys. Rev. Lett.}\ }\textbf {\bibinfo {volume} {110}},\
  \bibinfo {pages} {208001} (\bibinfo {year} {2013})}\BibitemShut {NoStop}%
\bibitem [{\citenamefont {Scholz}\ \emph {et~al.}(2018)\citenamefont {Scholz},
  \citenamefont {Jahanshahi}, \citenamefont {Ldov},\ and\ \citenamefont
  {L{\"{o}}wen}}]{Scholz2018}%
  \BibitemOpen
  \bibfield  {author} {\bibinfo {author} {\bibfnamefont {C.}~\bibnamefont
  {Scholz}}, \bibinfo {author} {\bibfnamefont {S.}~\bibnamefont {Jahanshahi}},
  \bibinfo {author} {\bibfnamefont {A.}~\bibnamefont {Ldov}}, \ and\ \bibinfo
  {author} {\bibfnamefont {H.}~\bibnamefont {L{\"{o}}wen}},\ }\href@noop {}
  {\bibfield  {journal} {\bibinfo  {journal} {Nat. Commun.}\ }\textbf {\bibinfo
  {volume} {9}},\ \bibinfo {pages} {5156} (\bibinfo {year} {2018})}\BibitemShut
  {NoStop}%
\bibitem [{\citenamefont {Bore}\ \emph {et~al.}(2016)\citenamefont {Bore},
  \citenamefont {Schindler}, \citenamefont {Lam}, \citenamefont {Bertin},\ and\
  \citenamefont {Dauchot}}]{Bore2016}%
  \BibitemOpen
  \bibfield  {author} {\bibinfo {author} {\bibfnamefont {S.~L.}\ \bibnamefont
  {Bore}}, \bibinfo {author} {\bibfnamefont {M.}~\bibnamefont {Schindler}},
  \bibinfo {author} {\bibfnamefont {K.-D. N.~T.}\ \bibnamefont {Lam}}, \bibinfo
  {author} {\bibfnamefont {E.}~\bibnamefont {Bertin}}, \ and\ \bibinfo {author}
  {\bibfnamefont {O.}~\bibnamefont {Dauchot}},\ }\href@noop {} {\bibfield
  {journal} {\bibinfo  {journal} {J. Stat. Mech.}\ }\textbf {\bibinfo {volume}
  {2016}},\ \bibinfo {pages} {033305} (\bibinfo {year} {2016})}\BibitemShut
  {NoStop}%
\bibitem [{\citenamefont {Casiulis}\ \emph {et~al.}(2020)\citenamefont
  {Casiulis}, \citenamefont {Tarzia}, \citenamefont {Cugliandolo},\ and\
  \citenamefont {Dauchot}}]{Casiulis2019b}%
  \BibitemOpen
  \bibfield  {author} {\bibinfo {author} {\bibfnamefont {M.}~\bibnamefont
  {Casiulis}}, \bibinfo {author} {\bibfnamefont {M.}~\bibnamefont {Tarzia}},
  \bibinfo {author} {\bibfnamefont {L.~F.}\ \bibnamefont {Cugliandolo}}, \ and\
  \bibinfo {author} {\bibfnamefont {O.}~\bibnamefont {Dauchot}},\ }\href@noop
  {} {\bibfield  {journal} {\bibinfo  {journal} {J. Stat. Mech.}\ }\textbf
  {\bibinfo {volume} {2020}},\ \bibinfo {pages} {013209} (\bibinfo {year}
  {2020})}\BibitemShut {NoStop}%
\bibitem [{\citenamefont {Casiulis}\ \emph {et~al.}(2019)\citenamefont
  {Casiulis}, \citenamefont {Tarzia}, \citenamefont {Cugliandolo},\ and\
  \citenamefont {Dauchot}}]{Casiulis2019}%
  \BibitemOpen
  \bibfield  {author} {\bibinfo {author} {\bibfnamefont {M.}~\bibnamefont
  {Casiulis}}, \bibinfo {author} {\bibfnamefont {M.}~\bibnamefont {Tarzia}},
  \bibinfo {author} {\bibfnamefont {L.~F.}\ \bibnamefont {Cugliandolo}}, \ and\
  \bibinfo {author} {\bibfnamefont {O.}~\bibnamefont {Dauchot}},\ }\href@noop
  {} {\bibfield  {journal} {\bibinfo  {journal} {J. Chem. Phys.}\ }\textbf
  {\bibinfo {volume} {150}},\ \bibinfo {pages} {154501} (\bibinfo {year}
  {2019})}\BibitemShut {NoStop}%
\bibitem [{Note1()}]{Note1}%
  \BibitemOpen
  \bibinfo {note} {See the SI for effect of $\tau $ on the shape of the
  correlations.}\BibitemShut {Stop}%
\bibitem [{\citenamefont {Cavagna}\ \emph {et~al.}(2008)\citenamefont
  {Cavagna}, \citenamefont {Cimarelli}, \citenamefont {Giardina}, \citenamefont
  {Orlandi}, \citenamefont {Parisi}, \citenamefont {Procaccini}, \citenamefont
  {Santagati},\ and\ \citenamefont {Stefanini}}]{Cavagna2008}%
  \BibitemOpen
  \bibfield  {author} {\bibinfo {author} {\bibfnamefont {A.}~\bibnamefont
  {Cavagna}}, \bibinfo {author} {\bibfnamefont {A.}~\bibnamefont {Cimarelli}},
  \bibinfo {author} {\bibfnamefont {I.}~\bibnamefont {Giardina}}, \bibinfo
  {author} {\bibfnamefont {A.}~\bibnamefont {Orlandi}}, \bibinfo {author}
  {\bibfnamefont {G.}~\bibnamefont {Parisi}}, \bibinfo {author} {\bibfnamefont
  {A.}~\bibnamefont {Procaccini}}, \bibinfo {author} {\bibfnamefont
  {R.}~\bibnamefont {Santagati}}, \ and\ \bibinfo {author} {\bibfnamefont
  {F.}~\bibnamefont {Stefanini}},\ }\href@noop {} {\bibfield  {journal}
  {\bibinfo  {journal} {Math. Biosci.}\ }\textbf {\bibinfo {volume} {214}},\
  \bibinfo {pages} {32} (\bibinfo {year} {2008})}\BibitemShut {NoStop}%
\bibitem [{\citenamefont {Ballerini}\ \emph {et~al.}(2008)\citenamefont
  {Ballerini}, \citenamefont {Cabibbo}, \citenamefont {Candelier},
  \citenamefont {Cavagna}, \citenamefont {Cisbani}, \citenamefont {Giardina},
  \citenamefont {Lecomte}, \citenamefont {Orlandi}, \citenamefont {Parisi},
  \citenamefont {Procaccini}, \citenamefont {Viale},\ and\ \citenamefont
  {Zdravkovic}}]{Ballerini2008a}%
  \BibitemOpen
  \bibfield  {author} {\bibinfo {author} {\bibfnamefont {M.}~\bibnamefont
  {Ballerini}}, \bibinfo {author} {\bibfnamefont {N.}~\bibnamefont {Cabibbo}},
  \bibinfo {author} {\bibfnamefont {R.}~\bibnamefont {Candelier}}, \bibinfo
  {author} {\bibfnamefont {A.}~\bibnamefont {Cavagna}}, \bibinfo {author}
  {\bibfnamefont {E.}~\bibnamefont {Cisbani}}, \bibinfo {author} {\bibfnamefont
  {I.}~\bibnamefont {Giardina}}, \bibinfo {author} {\bibfnamefont
  {V.}~\bibnamefont {Lecomte}}, \bibinfo {author} {\bibfnamefont
  {A.}~\bibnamefont {Orlandi}}, \bibinfo {author} {\bibfnamefont
  {G.}~\bibnamefont {Parisi}}, \bibinfo {author} {\bibfnamefont
  {A.}~\bibnamefont {Procaccini}}, \bibinfo {author} {\bibfnamefont
  {M.}~\bibnamefont {Viale}}, \ and\ \bibinfo {author} {\bibfnamefont
  {V.}~\bibnamefont {Zdravkovic}},\ }\href@noop {} {\bibfield  {journal}
  {\bibinfo  {journal} {Proceedings of the National Academy of Sciences of the
  United States of America}\ }\textbf {\bibinfo {volume} {105}},\ \bibinfo
  {pages} {1232} (\bibinfo {year} {2008})}\BibitemShut {NoStop}%
\bibitem [{\citenamefont {Halperin}\ and\ \citenamefont
  {Nelson}(1978)}]{Halperin1978}%
  \BibitemOpen
  \bibfield  {author} {\bibinfo {author} {\bibfnamefont {B.~I.}\ \bibnamefont
  {Halperin}}\ and\ \bibinfo {author} {\bibfnamefont {D.~R.}\ \bibnamefont
  {Nelson}},\ }\href@noop {} {\bibfield  {journal} {\bibinfo  {journal} {Phys.
  Rev. Lett.}\ }\textbf {\bibinfo {volume} {41}},\ \bibinfo {pages} {121}
  (\bibinfo {year} {1978})}\BibitemShut {NoStop}%
\bibitem [{\citenamefont {van Megen}\ \emph {et~al.}(1998)\citenamefont {van
  Megen}, \citenamefont {Mortensen}, \citenamefont {Williams},\ and\
  \citenamefont {M{\"{u}}ller}}]{VanMegen1998}%
  \BibitemOpen
  \bibfield  {author} {\bibinfo {author} {\bibfnamefont {W.}~\bibnamefont {van
  Megen}}, \bibinfo {author} {\bibfnamefont {T.~C.}\ \bibnamefont {Mortensen}},
  \bibinfo {author} {\bibfnamefont {S.~R.}\ \bibnamefont {Williams}}, \ and\
  \bibinfo {author} {\bibfnamefont {J.}~\bibnamefont {M{\"{u}}ller}},\
  }\href@noop {} {\bibfield  {journal} {\bibinfo  {journal} {Phys. Rev. E}\
  }\textbf {\bibinfo {volume} {58}},\ \bibinfo {pages} {6073} (\bibinfo {year}
  {1998})}\BibitemShut {NoStop}%
\bibitem [{\citenamefont {S{\'{a}}nchez-Miranda}\ \emph
  {et~al.}(2015)\citenamefont {S{\'{a}}nchez-Miranda}, \citenamefont
  {Bonilla-Capilla}, \citenamefont {Sarmien\-to G{\'{o}}\-mez}, \citenamefont
  {L{\'{a}}zaro-L{\'{a}}zaro}, \citenamefont {Ram{\'{i}}rez-Saito},
  \citenamefont {Medina-Noyola},\ and\ \citenamefont
  {Arauz-Lara}}]{Sanchez-Miranda2015}%
  \BibitemOpen
  \bibfield  {author} {\bibinfo {author} {\bibfnamefont {M.~J.}\ \bibnamefont
  {S{\'{a}}nchez-Miranda}}, \bibinfo {author} {\bibfnamefont {B.}~\bibnamefont
  {Bonilla-Capilla}}, \bibinfo {author} {\bibfnamefont {E.}~\bibnamefont
  {Sarmien\-to G{\'{o}}\-mez}}, \bibinfo {author} {\bibfnamefont
  {E.}~\bibnamefont {L{\'{a}}zaro-L{\'{a}}zaro}}, \bibinfo {author}
  {\bibfnamefont {A.}~\bibnamefont {Ram{\'{i}}rez-Saito}}, \bibinfo {author}
  {\bibfnamefont {M.}~\bibnamefont {Medina-Noyola}}, \ and\ \bibinfo {author}
  {\bibfnamefont {J.~L.}\ \bibnamefont {Arauz-Lara}},\ }\href@noop {}
  {\bibfield  {journal} {\bibinfo  {journal} {Soft Matter}\ }\textbf {\bibinfo
  {volume} {11}},\ \bibinfo {pages} {655} (\bibinfo {year} {2015})}\BibitemShut
  {NoStop}%
\bibitem [{\citenamefont {Tobochnik}\ and\ \citenamefont
  {Chester}(1979)}]{Tobochnik1979}%
  \BibitemOpen
  \bibfield  {author} {\bibinfo {author} {\bibfnamefont {J.}~\bibnamefont
  {Tobochnik}}\ and\ \bibinfo {author} {\bibfnamefont {G.~V.}\ \bibnamefont
  {Chester}},\ }\href@noop {} {\bibfield  {journal} {\bibinfo  {journal} {Phys.
  Rev. B}\ }\textbf {\bibinfo {volume} {20}},\ \bibinfo {pages} {3761}
  (\bibinfo {year} {1979})}\BibitemShut {NoStop}%
\bibitem [{\citenamefont {Sumino}\ \emph {et~al.}(2012)\citenamefont {Sumino},
  \citenamefont {Nagai}, \citenamefont {Shitaka}, \citenamefont {Tanaka},
  \citenamefont {Yoshikawa}, \citenamefont {Chat{\'{e}}},\ and\ \citenamefont
  {Oiwa}}]{Sumino2012}%
  \BibitemOpen
  \bibfield  {author} {\bibinfo {author} {\bibfnamefont {Y.}~\bibnamefont
  {Sumino}}, \bibinfo {author} {\bibfnamefont {K.~H.}\ \bibnamefont {Nagai}},
  \bibinfo {author} {\bibfnamefont {Y.}~\bibnamefont {Shitaka}}, \bibinfo
  {author} {\bibfnamefont {D.}~\bibnamefont {Tanaka}}, \bibinfo {author}
  {\bibfnamefont {K.}~\bibnamefont {Yoshikawa}}, \bibinfo {author}
  {\bibfnamefont {H.}~\bibnamefont {Chat{\'{e}}}}, \ and\ \bibinfo {author}
  {\bibfnamefont {K.}~\bibnamefont {Oiwa}},\ }\href@noop {} {\bibfield
  {journal} {\bibinfo  {journal} {Nature}\ }\textbf {\bibinfo {volume} {483}},\
  \bibinfo {pages} {448} (\bibinfo {year} {2012})}\BibitemShut {NoStop}%
\bibitem [{\citenamefont {Nagai}\ \emph {et~al.}(2015)\citenamefont {Nagai},
  \citenamefont {Sumino}, \citenamefont {Montagne}, \citenamefont {Aranson},\
  and\ \citenamefont {Chat{\'{e}}}}]{Nagai2015}%
  \BibitemOpen
  \bibfield  {author} {\bibinfo {author} {\bibfnamefont {K.~H.}\ \bibnamefont
  {Nagai}}, \bibinfo {author} {\bibfnamefont {Y.}~\bibnamefont {Sumino}},
  \bibinfo {author} {\bibfnamefont {R.}~\bibnamefont {Montagne}}, \bibinfo
  {author} {\bibfnamefont {I.~S.}\ \bibnamefont {Aranson}}, \ and\ \bibinfo
  {author} {\bibfnamefont {H.}~\bibnamefont {Chat{\'{e}}}},\ }\href@noop {}
  {\bibfield  {journal} {\bibinfo  {journal} {Phys. Rev. Lett.}\ }\textbf
  {\bibinfo {volume} {114}},\ \bibinfo {pages} {168001} (\bibinfo {year}
  {2015})}\BibitemShut {NoStop}%
\bibitem [{\citenamefont {Chen}\ \emph {et~al.}(2017)\citenamefont {Chen},
  \citenamefont {Liu}, \citenamefont {Shi}, \citenamefont {Chat{\'{e}}},\ and\
  \citenamefont {Wu}}]{Chen2017}%
  \BibitemOpen
  \bibfield  {author} {\bibinfo {author} {\bibfnamefont {C.}~\bibnamefont
  {Chen}}, \bibinfo {author} {\bibfnamefont {S.}~\bibnamefont {Liu}}, \bibinfo
  {author} {\bibfnamefont {X.~Q.}\ \bibnamefont {Shi}}, \bibinfo {author}
  {\bibfnamefont {H.}~\bibnamefont {Chat{\'{e}}}}, \ and\ \bibinfo {author}
  {\bibfnamefont {Y.}~\bibnamefont {Wu}},\ }\href@noop {} {\bibfield  {journal}
  {\bibinfo  {journal} {Nature}\ }\textbf {\bibinfo {volume} {542}},\ \bibinfo
  {pages} {210} (\bibinfo {year} {2017})}\BibitemShut {NoStop}%
\bibitem [{\citenamefont {Sugi}\ \emph {et~al.}(2019)\citenamefont {Sugi},
  \citenamefont {Ito}, \citenamefont {Nishimura},\ and\ \citenamefont
  {Nagai}}]{Sugi2019}%
  \BibitemOpen
  \bibfield  {author} {\bibinfo {author} {\bibfnamefont {T.}~\bibnamefont
  {Sugi}}, \bibinfo {author} {\bibfnamefont {H.}~\bibnamefont {Ito}}, \bibinfo
  {author} {\bibfnamefont {M.}~\bibnamefont {Nishimura}}, \ and\ \bibinfo
  {author} {\bibfnamefont {K.~H.}\ \bibnamefont {Nagai}},\ }\href@noop {}
  {\bibfield  {journal} {\bibinfo  {journal} {Nat. Commun.}\ }\textbf {\bibinfo
  {volume} {10}},\ \bibinfo {pages} {683} (\bibinfo {year} {2019})}\BibitemShut
  {NoStop}%
\bibitem [{\citenamefont {Burkhardt}(2007)}]{Burkhardt2007}%
  \BibitemOpen
  \bibfield  {author} {\bibinfo {author} {\bibfnamefont {T.~W.}\ \bibnamefont
  {Burkhardt}},\ }\href@noop {} {\bibfield  {journal} {\bibinfo  {journal} {J.
  Stat. Mech.}\ ,\ \bibinfo {pages} {P07004}} (\bibinfo {year}
  {2007})}\BibitemShut {NoStop}%
\bibitem [{\citenamefont {Kranz}\ and\ \citenamefont
  {Golestanian}(2019)}]{Kranz2019}%
  \BibitemOpen
  \bibfield  {author} {\bibinfo {author} {\bibfnamefont {W.~T.}\ \bibnamefont
  {Kranz}}\ and\ \bibinfo {author} {\bibfnamefont {R.}~\bibnamefont
  {Golestanian}},\ }\href@noop {} {\bibfield  {journal} {\bibinfo  {journal}
  {J. Chem. Phys.}\ }\textbf {\bibinfo {volume} {150}},\ \bibinfo {pages}
  {214111} (\bibinfo {year} {2019})}\BibitemShut {NoStop}%
\bibitem [{\citenamefont {Ghosh}\ \emph {et~al.}(2015)\citenamefont {Ghosh},
  \citenamefont {Li}, \citenamefont {Marchegiani},\ and\ \citenamefont
  {Marchesoni}}]{Ghosh2015}%
  \BibitemOpen
  \bibfield  {author} {\bibinfo {author} {\bibfnamefont {P.~K.}\ \bibnamefont
  {Ghosh}}, \bibinfo {author} {\bibfnamefont {Y.}~\bibnamefont {Li}}, \bibinfo
  {author} {\bibfnamefont {G.}~\bibnamefont {Marchegiani}}, \ and\ \bibinfo
  {author} {\bibfnamefont {F.}~\bibnamefont {Marchesoni}},\ }\href@noop {}
  {\bibfield  {journal} {\bibinfo  {journal} {J. Chem. Phys.}\ }\textbf
  {\bibinfo {volume} {143}},\ \bibinfo {pages} {211101} (\bibinfo {year}
  {2015})}\BibitemShut {NoStop}%
\bibitem [{\citenamefont {Fily}\ and\ \citenamefont
  {Marchetti}(2012)}]{Fily2012}%
  \BibitemOpen
  \bibfield  {author} {\bibinfo {author} {\bibfnamefont {Y.}~\bibnamefont
  {Fily}}\ and\ \bibinfo {author} {\bibfnamefont {M.~C.}\ \bibnamefont
  {Marchetti}},\ }\href@noop {} {\bibfield  {journal} {\bibinfo  {journal}
  {Phys. Rev. Lett.}\ }\textbf {\bibinfo {volume} {108}},\ \bibinfo {pages}
  {235702} (\bibinfo {year} {2012})}\BibitemShut {NoStop}%
\bibitem [{\citenamefont {Winkler}\ \emph {et~al.}(2015)\citenamefont
  {Winkler}, \citenamefont {Wysocki},\ and\ \citenamefont
  {Gompper}}]{Winkler2015}%
  \BibitemOpen
  \bibfield  {author} {\bibinfo {author} {\bibfnamefont {R.~G.}\ \bibnamefont
  {Winkler}}, \bibinfo {author} {\bibfnamefont {A.}~\bibnamefont {Wysocki}}, \
  and\ \bibinfo {author} {\bibfnamefont {G.}~\bibnamefont {Gompper}},\
  }\href@noop {} {\bibfield  {journal} {\bibinfo  {journal} {Soft Matter}\
  }\textbf {\bibinfo {volume} {11}},\ \bibinfo {pages} {6680} (\bibinfo {year}
  {2015})}\BibitemShut {NoStop}%
\bibitem [{\citenamefont {Cavagna}\ \emph
  {et~al.}(2019{\natexlab{b}})\citenamefont {Cavagna}, \citenamefont
  {Giardina},\ and\ \citenamefont {Viale}}]{CavagnaComment}%
  \BibitemOpen
  \bibfield  {author} {\bibinfo {author} {\bibfnamefont {A.}~\bibnamefont
  {Cavagna}}, \bibinfo {author} {\bibfnamefont {I.}~\bibnamefont {Giardina}}, \
  and\ \bibinfo {author} {\bibfnamefont {M.}~\bibnamefont {Viale}},\
  }\href@noop {} {\bibfield  {journal} {\bibinfo  {journal} {Arxiv Prepr.}\ ,\
  \bibinfo {pages} {1912.07056v1}} (\bibinfo {year}
  {2019}{\natexlab{b}})}\BibitemShut {NoStop}%
\bibitem [{\citenamefont {Ginot}\ \emph {et~al.}(2018)\citenamefont {Ginot},
  \citenamefont {Theurkauff}, \citenamefont {Detcheverry}, \citenamefont
  {Ybert},\ and\ \citenamefont {Cottin-Bizonne}}]{Ginot2018}%
  \BibitemOpen
  \bibfield  {author} {\bibinfo {author} {\bibfnamefont {F.}~\bibnamefont
  {Ginot}}, \bibinfo {author} {\bibfnamefont {I.}~\bibnamefont {Theurkauff}},
  \bibinfo {author} {\bibfnamefont {F.}~\bibnamefont {Detcheverry}}, \bibinfo
  {author} {\bibfnamefont {C.}~\bibnamefont {Ybert}}, \ and\ \bibinfo {author}
  {\bibfnamefont {C.}~\bibnamefont {Cottin-Bizonne}},\ }\href@noop {}
  {\bibfield  {journal} {\bibinfo  {journal} {Nat. Commun.}\ }\textbf {\bibinfo
  {volume} {9}},\ \bibinfo {pages} {696} (\bibinfo {year} {2018})}\BibitemShut
  {NoStop}%
\bibitem [{\citenamefont {van~der Linden}\ \emph {et~al.}(2019)\citenamefont
  {van~der Linden}, \citenamefont {Alexander}, \citenamefont {Aarts},\ and\
  \citenamefont {Dauchot}}]{VanderLinden2019}%
  \BibitemOpen
  \bibfield  {author} {\bibinfo {author} {\bibfnamefont {M.~N.}\ \bibnamefont
  {van~der Linden}}, \bibinfo {author} {\bibfnamefont {L.~C.}\ \bibnamefont
  {Alexander}}, \bibinfo {author} {\bibfnamefont {D.~G. A.~L.}\ \bibnamefont
  {Aarts}}, \ and\ \bibinfo {author} {\bibfnamefont {O.}~\bibnamefont
  {Dauchot}},\ }\href@noop {} {\bibfield  {journal} {\bibinfo  {journal} {Phys.
  Rev. Lett.}\ }\textbf {\bibinfo {volume} {123}},\ \bibinfo {pages} {098001}
  (\bibinfo {year} {2019})}\BibitemShut {NoStop}%
\bibitem [{\citenamefont {Petrelli}\ \emph {et~al.}(2018)\citenamefont
  {Petrelli}, \citenamefont {Digregorio}, \citenamefont {Cugliandolo},
  \citenamefont {Gonnella},\ and\ \citenamefont {Suma}}]{Petrelli2018}%
  \BibitemOpen
  \bibfield  {author} {\bibinfo {author} {\bibfnamefont {I.}~\bibnamefont
  {Petrelli}}, \bibinfo {author} {\bibfnamefont {P.}~\bibnamefont
  {Digregorio}}, \bibinfo {author} {\bibfnamefont {L.~F.}\ \bibnamefont
  {Cugliandolo}}, \bibinfo {author} {\bibfnamefont {G.}~\bibnamefont
  {Gonnella}}, \ and\ \bibinfo {author} {\bibfnamefont {A.}~\bibnamefont
  {Suma}},\ }\href@noop {} {\bibfield  {journal} {\bibinfo  {journal} {Eur.
  Phys. J. E}\ }\textbf {\bibinfo {volume} {41}},\ \bibinfo {pages} {128}
  (\bibinfo {year} {2018})}\BibitemShut {NoStop}%
\end{thebibliography}%


\begin{thebibliography}{10}%
\makeatletter
\providecommand \@ifxundefined [1]{%
 \@ifx{#1\undefined}
}%
\providecommand \@ifnum [1]{%
 \ifnum #1\expandafter \@firstoftwo
 \else \expandafter \@secondoftwo
 \fi
}%
\providecommand \@ifx [1]{%
 \ifx #1\expandafter \@firstoftwo
 \else \expandafter \@secondoftwo
 \fi
}%
\providecommand \natexlab [1]{#1}%
\providecommand \enquote  [1]{``#1''}%
\providecommand \bibnamefont  [1]{#1}%
\providecommand \bibfnamefont [1]{#1}%
\providecommand \citenamefont [1]{#1}%
\providecommand \href@noop [0]{\@secondoftwo}%
\providecommand \href [0]{\begingroup \@sanitize@url \@href}%
\providecommand \@href[1]{\@@startlink{#1}\@@href}%
\providecommand \@@href[1]{\endgroup#1\@@endlink}%
\providecommand \@sanitize@url [0]{\catcode `\\12\catcode `\$12\catcode
  `\&12\catcode `\#12\catcode `\^12\catcode `\_12\catcode `\%12\relax}%
\providecommand \@@startlink[1]{}%
\providecommand \@@endlink[0]{}%
\providecommand \url  [0]{\begingroup\@sanitize@url \@url }%
\providecommand \@url [1]{\endgroup\@href {#1}{\urlprefix }}%
\providecommand \urlprefix  [0]{URL }%
\providecommand \Eprint [0]{\href }%
\providecommand \doibase [0]{http://dx.doi.org/}%
\providecommand \selectlanguage [0]{\@gobble}%
\providecommand \bibinfo  [0]{\@secondoftwo}%
\providecommand \bibfield  [0]{\@secondoftwo}%
\providecommand \translation [1]{[#1]}%
\providecommand \BibitemOpen [0]{}%
\providecommand \bibitemStop [0]{}%
\providecommand \bibitemNoStop [0]{.\EOS\space}%
\providecommand \EOS [0]{\spacefactor3000\relax}%
\providecommand \BibitemShut  [1]{\csname bibitem#1\endcsname}%
\let\auto@bib@innerbib\@empty
\bibitem [{\citenamefont {Bore}\ \emph {et~al.}(2016)\citenamefont {Bore},
  \citenamefont {Schindler}, \citenamefont {Lam}, \citenamefont {Bertin},\ and\
  \citenamefont {Dauchot}}]{Bore2016}%
  \BibitemOpen
  \bibfield  {author} {\bibinfo {author} {\bibfnamefont {S.~L.}\ \bibnamefont
  {Bore}}, \bibinfo {author} {\bibfnamefont {M.}~\bibnamefont {Schindler}},
  \bibinfo {author} {\bibfnamefont {K.-D. N.~T.}\ \bibnamefont {Lam}}, \bibinfo
  {author} {\bibfnamefont {E.}~\bibnamefont {Bertin}}, \ and\ \bibinfo {author}
  {\bibfnamefont {O.}~\bibnamefont {Dauchot}},\ }\href@noop {} {\bibfield
  {journal} {\bibinfo  {journal} {J. Stat. Mech.}\ }\textbf {\bibinfo {volume}
  {2016}},\ \bibinfo {pages} {033305} (\bibinfo {year} {2016})}\BibitemShut
  {NoStop}%
\bibitem [{\citenamefont {Casiulis}\ \emph {et~al.}(2020)\citenamefont
  {Casiulis}, \citenamefont {Tarzia}, \citenamefont {Cugliandolo},\ and\
  \citenamefont {Dauchot}}]{Casiulis2019b}%
  \BibitemOpen
  \bibfield  {author} {\bibinfo {author} {\bibfnamefont {M.}~\bibnamefont
  {Casiulis}}, \bibinfo {author} {\bibfnamefont {M.}~\bibnamefont {Tarzia}},
  \bibinfo {author} {\bibfnamefont {L.~F.}\ \bibnamefont {Cugliandolo}}, \ and\
  \bibinfo {author} {\bibfnamefont {O.}~\bibnamefont {Dauchot}},\ }\href@noop
  {} {\bibfield  {journal} {\bibinfo  {journal} {J. Stat. Mech.}\ }\textbf
  {\bibinfo {volume} {2020}},\ \bibinfo {pages} {013209} (\bibinfo {year}
  {2020})}\BibitemShut {NoStop}%
\bibitem [{\citenamefont {Casiulis}\ \emph {et~al.}(2019)\citenamefont
  {Casiulis}, \citenamefont {Tarzia}, \citenamefont {Cugliandolo},\ and\
  \citenamefont {Dauchot}}]{Casiulis2019}%
  \BibitemOpen
  \bibfield  {author} {\bibinfo {author} {\bibfnamefont {M.}~\bibnamefont
  {Casiulis}}, \bibinfo {author} {\bibfnamefont {M.}~\bibnamefont {Tarzia}},
  \bibinfo {author} {\bibfnamefont {L.~F.}\ \bibnamefont {Cugliandolo}}, \ and\
  \bibinfo {author} {\bibfnamefont {O.}~\bibnamefont {Dauchot}},\ }\href@noop
  {} {\bibfield  {journal} {\bibinfo  {journal} {J. Chem. Phys.}\ }\textbf
  {\bibinfo {volume} {150}},\ \bibinfo {pages} {154501} (\bibinfo {year}
  {2019})}\BibitemShut {NoStop}%
\bibitem [{\citenamefont {Cavagna}\ \emph {et~al.}(2010)\citenamefont
  {Cavagna}, \citenamefont {Cimarelli}, \citenamefont {Giardina}, \citenamefont
  {Parisi}, \citenamefont {Santagati}, \citenamefont {Stefanini},\ and\
  \citenamefont {Viale}}]{Cavagna2010}%
  \BibitemOpen
  \bibfield  {author} {\bibinfo {author} {\bibfnamefont {A.}~\bibnamefont
  {Cavagna}}, \bibinfo {author} {\bibfnamefont {A.}~\bibnamefont {Cimarelli}},
  \bibinfo {author} {\bibfnamefont {I.}~\bibnamefont {Giardina}}, \bibinfo
  {author} {\bibfnamefont {G.}~\bibnamefont {Parisi}}, \bibinfo {author}
  {\bibfnamefont {R.}~\bibnamefont {Santagati}}, \bibinfo {author}
  {\bibfnamefont {F.}~\bibnamefont {Stefanini}}, \ and\ \bibinfo {author}
  {\bibfnamefont {M.}~\bibnamefont {Viale}},\ }\href@noop {} {\bibfield
  {journal} {\bibinfo  {journal} {Proc. Natl. Acad. Sci.}\ }\textbf {\bibinfo
  {volume} {107}},\ \bibinfo {pages} {11865} (\bibinfo {year}
  {2010})}\BibitemShut {NoStop}%
\bibitem [{\citenamefont {Gel'Fand}\ and\ \citenamefont
  {Shilov}(1968)}]{GelFand1968}%
  \BibitemOpen
  \bibfield  {author} {\bibinfo {author} {\bibfnamefont {I.}~\bibnamefont
  {Gel'Fand}}\ and\ \bibinfo {author} {\bibfnamefont {G.}~\bibnamefont
  {Shilov}},\ }\href@noop {} {\emph {\bibinfo {title} {{Generalized Functions,
  Vol. 1}}}}\ (\bibinfo  {publisher} {Academic Press},\ \bibinfo {year}
  {1968})\BibitemShut {NoStop}%
\bibitem [{\citenamefont {Abramowitz}\ and\ \citenamefont
  {Stegun}(1972)}]{Abramowitz1972}%
  \BibitemOpen
  \bibfield  {author} {\bibinfo {author} {\bibfnamefont {M.}~\bibnamefont
  {Abramowitz}}\ and\ \bibinfo {author} {\bibfnamefont {I.~A.}\ \bibnamefont
  {Stegun}},\ }\href@noop {} {\emph {\bibinfo {title} {{Handbook of
  Mathematical Functions}}}}\ (\bibinfo  {publisher} {National Bureau of
  Standards, Applied Mathematics Series},\ \bibinfo {year} {1972})\BibitemShut
  {NoStop}%
\bibitem [{\citenamefont {Landau}\ and\ \citenamefont
  {Lifshitz}(1986)}]{Landau1986}%
  \BibitemOpen
  \bibfield  {author} {\bibinfo {author} {\bibfnamefont {L.~D.}\ \bibnamefont
  {Landau}}\ and\ \bibinfo {author} {\bibfnamefont {E.~M.}\ \bibnamefont
  {Lifshitz}},\ }\href@noop {} {\emph {\bibinfo {title} {{Course of Theoretical
  Physics Vol. 7: Theory of Elasticity}}}},\ \bibinfo {edition} {3rd}\ ed.\
  (\bibinfo  {publisher} {Pergamon Press},\ \bibinfo {year} {1986})\BibitemShut
  {NoStop}%
\bibitem [{\citenamefont {Attanasi}\ \emph {et~al.}(2015)\citenamefont
  {Attanasi}, \citenamefont {Cavagna}, \citenamefont {{Del Castello}},
  \citenamefont {Giardina}, \citenamefont {Jelic}, \citenamefont {Melillo},
  \citenamefont {Parisi}, \citenamefont {Pohl}, \citenamefont {Shen},\ and\
  \citenamefont {Viale}}]{Attanasi2015}%
  \BibitemOpen
  \bibfield  {author} {\bibinfo {author} {\bibfnamefont {A.}~\bibnamefont
  {Attanasi}}, \bibinfo {author} {\bibfnamefont {A.}~\bibnamefont {Cavagna}},
  \bibinfo {author} {\bibfnamefont {L.}~\bibnamefont {{Del Castello}}},
  \bibinfo {author} {\bibfnamefont {I.}~\bibnamefont {Giardina}}, \bibinfo
  {author} {\bibfnamefont {A.}~\bibnamefont {Jelic}}, \bibinfo {author}
  {\bibfnamefont {S.}~\bibnamefont {Melillo}}, \bibinfo {author} {\bibfnamefont
  {L.}~\bibnamefont {Parisi}}, \bibinfo {author} {\bibfnamefont
  {O.}~\bibnamefont {Pohl}}, \bibinfo {author} {\bibfnamefont {E.}~\bibnamefont
  {Shen}}, \ and\ \bibinfo {author} {\bibfnamefont {M.}~\bibnamefont {Viale}},\
  }\href@noop {} {\bibfield  {journal} {\bibinfo  {journal} {J. R. Soc.
  Interface}\ }\textbf {\bibinfo {volume} {12}},\ \bibinfo {pages} {20150319}
  (\bibinfo {year} {2015})}\BibitemShut {NoStop}%
\bibitem [{\citenamefont {Cavagna}\ \emph {et~al.}(2018)\citenamefont
  {Cavagna}, \citenamefont {Giardina},\ and\ \citenamefont
  {Grigera}}]{Cavagna2018}%
  \BibitemOpen
  \bibfield  {author} {\bibinfo {author} {\bibfnamefont {A.}~\bibnamefont
  {Cavagna}}, \bibinfo {author} {\bibfnamefont {I.}~\bibnamefont {Giardina}}, \
  and\ \bibinfo {author} {\bibfnamefont {T.~S.}\ \bibnamefont {Grigera}},\
  }\href@noop {} {\bibfield  {journal} {\bibinfo  {journal} {Phys. Rep.}\
  }\textbf {\bibinfo {volume} {728}},\ \bibinfo {pages} {1} (\bibinfo {year}
  {2018})}\BibitemShut {NoStop}%
\bibitem [{\citenamefont {Attanasi}\ \emph {et~al.}(2014)\citenamefont
  {Attanasi}, \citenamefont {Cavagna}, \citenamefont {{Del Castello}},
  \citenamefont {Giardina}, \citenamefont {Grigera}, \citenamefont {Jelic},
  \citenamefont {Melillo}, \citenamefont {Parisi}, \citenamefont {Pohl},
  \citenamefont {Shen},\ and\ \citenamefont {Viale}}]{Attanasi2014a}%
  \BibitemOpen
  \bibfield  {author} {\bibinfo {author} {\bibfnamefont {A.}~\bibnamefont
  {Attanasi}}, \bibinfo {author} {\bibfnamefont {A.}~\bibnamefont {Cavagna}},
  \bibinfo {author} {\bibfnamefont {L.}~\bibnamefont {{Del Castello}}},
  \bibinfo {author} {\bibfnamefont {I.}~\bibnamefont {Giardina}}, \bibinfo
  {author} {\bibfnamefont {T.~S.}\ \bibnamefont {Grigera}}, \bibinfo {author}
  {\bibfnamefont {A.}~\bibnamefont {Jelic}}, \bibinfo {author} {\bibfnamefont
  {S.}~\bibnamefont {Melillo}}, \bibinfo {author} {\bibfnamefont
  {L.}~\bibnamefont {Parisi}}, \bibinfo {author} {\bibfnamefont
  {O.}~\bibnamefont {Pohl}}, \bibinfo {author} {\bibfnamefont {E.}~\bibnamefont
  {Shen}}, \ and\ \bibinfo {author} {\bibfnamefont {M.}~\bibnamefont {Viale}},\
  }\href@noop {} {\bibfield  {journal} {\bibinfo  {journal} {Nat. Phys.}\
  }\textbf {\bibinfo {volume} {10}},\ \bibinfo {pages} {691} (\bibinfo {year}
  {2014})}\BibitemShut {NoStop}%
\end{thebibliography}%

\end{document}


\preprint{APS/123-QED}

\title{Supplementary Information for \\
``Velocity and Speed Correlations in Hamiltonian Flocks"} 

\author{Mathias Casiulis}
\affiliation{Sorbonne Universit\'{e}, CNRS UMR 7600, Laboratoire de Physique Th\'{e}orique de la Matière Condens\'{e}e, LPTMC, 4 place Jussieu, Couloir 12-13, 5ème \'{e}tage, 75252 Paris Cedex 05, France}
\email{casiulis@lptmc.jussieu.fr}
\author{Marco Tarzia}
\affiliation{Sorbonne Universit\'{e}, CNRS UMR 7600, Laboratoire de Physique Th\'{e}orique de la Matière Condens\'{e}e, LPTMC, 4 place Jussieu, Couloir 12-13, 5ème \'{e}tage, 75252 Paris Cedex 05, France}
\affiliation{Institut Universitaire de France, 1 rue Descartes, 75231 Paris Cedex 05, France}
\author{Leticia F. Cugliandolo}
\affiliation{Sorbonne Universit\'{e}, CNRS UMR 7589, Laboratoire de Physique Th\'{e}orique et Hautes Energies, LPTHE, 4 place Jussieu, Couloir 13-14, 5ème \'{e}tage, 75252 Paris Cedex 05, France}
\affiliation{Institut Universitaire de France, 1 rue Descartes, 75231 Paris Cedex 05, France}
\author{Olivier Dauchot}
\affiliation{UMR Gulliver 7083 CNRS, ESPCI Paris, PSL Research University, 10 rue Vauquelin, 75005 Paris, France}

\date{\today}

\maketitle

\section{Movie captions\label{sec:Mov}}
\textit{Movie1.mp4 --} 
Consecutive snapshots of a system of $N = 512$ particles, with $K = 0.03$, at a packing fraction $\phi \approx 0.15$, as obtained from Molecular Dynamics at 
temperature $ T \approx 1\cdot 10^{-2}$.
Consecutive snapshots are separated by a time delay $\Delta t = 50$ in the rescaled units defined in Sec.~\ref{sec:Lagrangian}.
Particles are color-coded according to the direction of their spin, and a large black arrow represents the total magnetization of the system.
The periodic boundary conditions were unwound, so that net displacements are easier to see.

\textit{Movie2.mp4 --} 
Same as in the previous movie, except this time the snapshots are centered on the center of mass of the system, so that only 
rotational and deformational motions are represented.
The total magnetization here represented on a tagged particle to improve readability.



\section{From the Lagrangian to the Hamiltonian description}
\label{sec:Lagrangian}

In the main text we discuss the properties of a Hamiltonian system derived from its original definition in the 
Lagrangian formalism:~\cite{Bore2016,Casiulis2019b}
\begin{align}
 \mathcal{L} &= \sum\limits_{i = 1}^{N} \frac{m}{2} \dot{\bm{r}}_i^2 + 
                \sum\limits_{i = 1}^{N}\frac{I}{2} {\dot{\bm{s}}_i}^2 +
                \sum\limits_{i = 1}^{N} K \dot{\bm{r}}_i \cdot \bm{s}_i - \frac{U_0}{2} \sum\limits_{k \neq i} U(r_{ik}) 
  + \frac{J_0}{2} \sum\limits_{k \neq i} J(r_{ik}) \cos\theta_{ik}
  \; .
  \label{eq:Lagrangian}
\end{align}
The model describes the dynamics of $N$ interacting particles confined to move in a $2d$ periodic square box of linear size $L_0$. The 
particles carry continuous planar spins with unit modulus. 
We call $\bm{r}_i$ the position of the $i$th particle, and $\dot{\bm{r}}_i$ its velocity.
We note $r_{ik}= |{\bm r}_i - {\bm r}_k|$ the distance between the centers of the particles $i$ and $k$.
$\theta_i$ is the angle that the spin forms with a 
reference axis and fully parametrizes the continuous $2d$ spin $\bm{s}_i$ of unit modulus. The time derivative of the spin vector is indicated by 
$\dot{\bm{s}}_i$. 
Finally, $\theta_{ik}$ is the angle between the spin of particle $i$ and the one of particle $k$. 
The model has several parameters and two potentials. 
$m$ is the mass of each particle and $I$ its moment of inertia.
$U$ is a short-ranged, isotropic and purely repulsive two-body interaction potential and 
$J$ is a short-ranged and isotropic ferromagnetic coupling between the spins and we have extracted their 
strengths as the parameters $U_0$ and $J_0$. Following the choices of 
Refs.~[\onlinecite{Bore2016,Casiulis2019,Casiulis2019b}] 
we use the forms
\begin{eqnarray}
 J(r) &=&     (\sigma - r)^2 \, \Theta(\sigma - r)  \; , \nonumber \\
 U(r) &=& (\sigma - r)^4 \, \Theta(\sigma - r) \; ,  \label{eq:Int}
\end{eqnarray}
where $\Theta$ is a Heaviside step function.
These potentials ensure well-behaved discretized Hamiltonian dynamics since they smoothly decrease to zero.
$\sigma$ is a range that we fix to $1$, resulting in an effective hard radius $r_0 = \sigma/4=1/4$ at zero temperature.
We define the packing fraction as $\phi \equiv \pi N r_0^2 / L^2$.
We fixed $J_0=1$ and $U_0 = 4$ so that both potentials are equal at half-range. 
Finally, $K$ is the parameter that controls the strength of the spin-velocity coupling, 
the term that breaks the Galilean invariance and leads to collective motion.

In the main text, we scale the dynamic variables, space-time variables, and parameters according to the transformations $\bm{r}/\sqrt{I/m} \to \bm{r}$, $t/\sqrt{I/J_0} \to t$, $K/\sqrt{m J_0} \to K$, $L/J_0 \to L$, $U_0/J_0 \to U_0$, and we absorb $U_0$ in the definition of the two-body potential $U$.
In the Hamiltonian formalism the Lagrangian~(\ref{eq:Lagrangian}) transforms into
\begin{align}
   \mathcal{H} &= \sum\limits_{i = 1}^{N} \frac{1}{2} \bm{p}_i^2 + 
                \sum\limits_{i = 1}^{N}\frac{1}{2} \omega_i^2 -
                \sum\limits_{i = 1}^{N} K \bm{p}_i \cdot \bm{s}_i + \frac{1}{2} \sum\limits_{k \neq i} U(r_{ik}) - \frac{1}{2} \sum\limits_{k \neq i} J(r_{ik}) \cos\theta_{ik} \label{eq:Ham}
  \; ,
\end{align}
where we defined the canonical momenta $\omega_i = \dot{\theta_i}$ and $\bm{p}_i = \dot{\bm{r}} + K\bm{s}_i$.

\section{Numerical methods\label{sec:Num}}

We simulate the dynamics starting from random states with uniformly distributed $\left\{\bm{r}_i, \theta_i \right\}_{i=1..N}$ and $\left\{\bm{p}_i, \omega_i \right\}_{i=1..N}$ drawn from centered, reduced Gaussian distributions. 
Such initial states were placed into a square box with periodic boundary conditions and, after giving some time for the dynamics to settle in, are subjected  to a numerical annealing to prepare colder equilibrated states. 
Numerical annealings are performed by multiplying all rotational velocities by $\lambda_{A} = 0.9999$ every 100 time units in our adimensionalized variable, with an integration time step equal to $\delta t = 10^{-3}$ in the same units. 
This method enables us to reach low-energy states which, if the cooling is slow enough, should be equilibrium states of the system.
Mean values are computed over $10^2-10^3$ independent configurations.

Due to the periodic boundary conditions, the angular momentum of the system is, generically, not conserved.
In particular, a homogeneous fluid that fills the periodic simulation box cannot rotate around a point due to this choice of boundary conditions.
However, the conservation of angular momentum is recovered when considering very cold droplets that are much smaller than the simulation box and are surrounded only by vacuum (such as in the case of Fig.~4$(c)-(d)$ of the main text).
Indeed, in these cases, the droplets see no difference between moving in vacuum with periodic boundary conditions or in free space.
Therefore, if one ``unwraps'' the periodic boundary conditions, and maps it onto free space by counting how many times the droplet has left the simulation box in each direction, the angular momentum in these new coordinates is conserved (within numerical accuracy).
Furthermore, as discussed in the main text, for the specific choice $\bm{P} = \bm{0}$, the angular momentum of the system does not depend on the position of its center of mass, but only on the positions of the particles relative to the center of mass of the droplet.
We checked in Fig. 4$(c)$ of the main text that the components of the angular momentum with respect to the center of mass do indeed compensate for a cold droplet.

In the main text, similarly, we compute the displacements $\bm{u}$ by unwrapping the periodic boundary condition, so as to measure the actual distances travelled by the particles of the fluid.
More precisely, if a particle moves from a coordinate $L_0 - x_1$ to a coordinate $x_2$, with $0 < x_1, x_2 < L_0$, and $L_0$ the linear size of the simulation box, 
the associated displacement we compute is equal to $x_2 + x_1 + (k - 1) L_0$, with $k$ the number of times that the particle has crossed a boundary of the box, counted positively to the right and negatively to the left.

\section{Density-Temperature Phase Diagram\label{sec:PhD}}


We showed in Ref.~[\onlinecite{Casiulis2019}] that, in the case $K = 0$, the model described by the Hamiltonian~(\ref{eq:Ham}) features a phase separation between a ferromagnetic liquid and a paramagnetic gas at low densities and temperatures.
This phase separation, which is observed even when the pairwise potential $U(r)$ is purely repulsive, is induced by the ferromagnetic coupling $J(r)$.
Indeed, a $J(r)$ that decreases as $r$ increases creates an effective attraction between aligned spins, that is sufficient to cause phase separation, 
that we coined Ferromagnetism-Induced Phase Separation (FIPS)~\cite{Casiulis2019}.
Because $J$ is the only source of attraction in the system, we also showed that the left-most spinodal of the liquid-gas phase separation was the lower part of the finite-size Curie line, so that a tricritical point sits at the top of the coexistence line, which presents a cusp.
\begin{figure}
    \centering
    \includegraphics[width = .40\columnwidth]{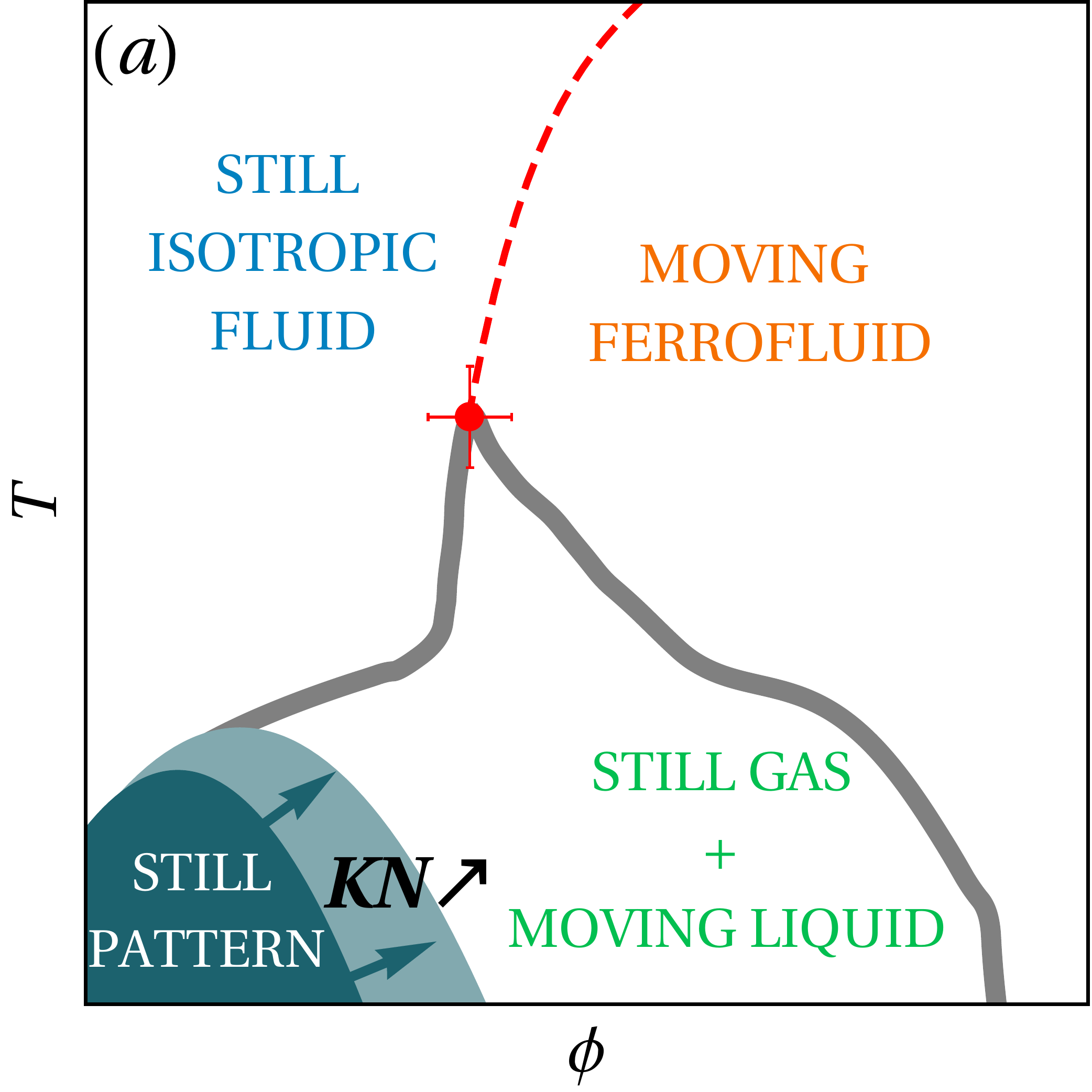}
    \includegraphics[height=.40\columnwidth]{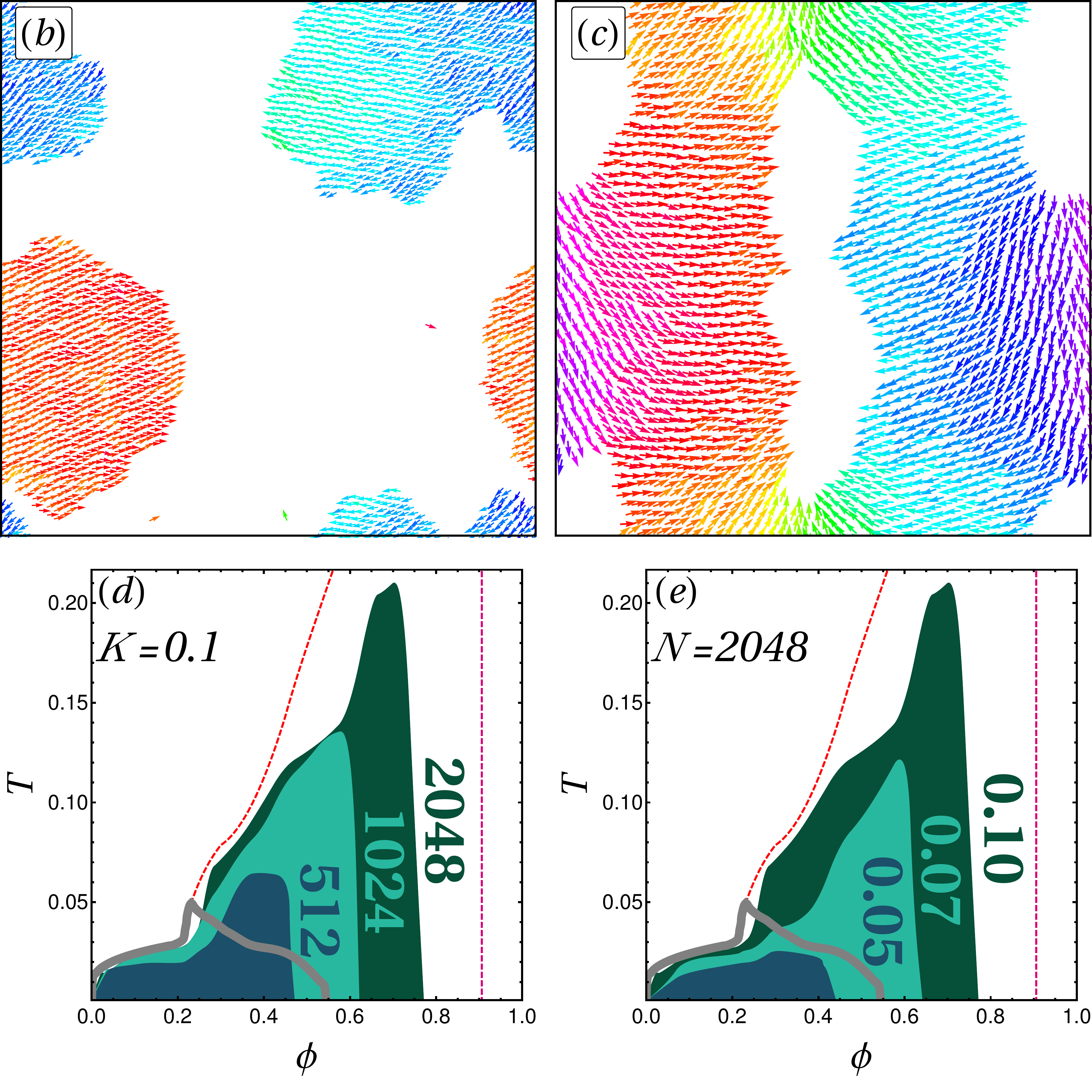}
    \caption{\textbf{Density-Temperature phase diagram.}
     $(a)$ Sketch of the finite-size phase diagram in the packing fraction, temperature plane, for a low value of $K N$. Increasing the value of $K N$ bloats the patterned region with a small $m$.
    $(b)-(c)$ Examples of spin maps of low-temperature ($T\approx 0.01$) patterned phase-separated states: 2 anti-magnetized droplets obtained for $N = 2048, K = 0.05, \phi = 0.20$, and opened up vortices found for $N = 2048, K = 0.05, \phi = 0.40$. 
    Spins are colored depending on their orientation.
    $(d)-(e)$ Extent of the patterned regions, as obtained from simulations when varying $(d)$ $N$ at a fixed $K = 0.1$, or $(e)$ $K$ at a fixed $N = 2048$.
    The boundaries of the patterned states are obtained by visual inspection of states obtained by slow annealings, for 16 different values of the packing fraction in the interval $0< \phi \leq 0.8$.}
    \label{fig:LowDens}
\end{figure}

For $K > 0$, a phase separation between a liquid and a gas is still expected at low densities and temperatures, since the spin-mediated attraction between particles is not affected by the presence of a spin-velocity coupling.
Furthermore, we showed in Ref.~[\onlinecite{Casiulis2019b}] that increasing $K$ or $N$ in homogeneous states increases the cost associated to the kinetic energy.
As a result, there is a crossover from polar, moving states to states that contain patterns in the magnetization field such that $m = v_G = 0$ as $T\to 0$, like vortices or solitons, thereby avoiding the kinetic energy cost of a global velocity.
This mechanism should also apply to phase-separated domains, as it only relies on energetic arguments, so that we expect phase-separated, still, patterned states.
It is \textit{a priori} hard to predict all the possible patterns of such states.
However, increasing the density at a constant temperature, and for given values of $K$ and $N$, one intuitively expects 
 the mean number of neighbours of each particle to increase.
As a result, curving the magnetization field should become more and more costly as the density increases.
Therefore, one expects a domain of still, patterned states at low densities and temperatures, that grows as $K$ or $N$ is increased.
This picture is schematically drawn atop the coexistence line and Curie line from Ref.~[\onlinecite{Casiulis2019}] in Fig.~\ref{fig:LowDens}$(a)$.

In order to check this picture, we simulate systems at various packing fractions, and for a few sets of values of $K$ and $N$.
At low densities, we do observe a rather large zoology of patterned phase-separated states, two examples of which are shown in Fig.~\ref{fig:LowDens}$(b)-(c)$.
At very low densities (close to the left-most side of the coexistence line), as illustrated by Fig.~\ref{fig:LowDens}$(b)$, the system can organize into two anti-magnetized droplets, and thus replace the energetic cost linked to curving the magnetization field by a surface tension cost.
At higher densities, closer to the right-most side of the coexistence line, patterns reminiscent of those observed in the homogeneous case are observed, as illustrated by Fig.~\ref{fig:LowDens}$(c)$.
This configuration resembles the vortex configurations described in the homogeneous case in Ref.~[\onlinecite{Casiulis2019b}], but where the system decreased the cost of the magnetic pattern by  ``unzipping'' the line between two of the vortices, thereby creating a hole with no topological charge.
The holes at the cores of the two remaining topological defects are also enlarged, as ill-aligned spins are more repulsive.
We also report in Fig.~\ref{fig:LowDens}$(d), (e)$ the location of such patterned states when varying $N$ at a fixed $K$ (Fig.~\ref{fig:LowDens}$(d)$), and vice-versa (Fig.~\ref{fig:LowDens}$(e)$).
These results confirm the picture that increasing $K$ or $N$ causes the patterned region to invade the ferromagnetic domains of the phase diagram, starting from the bottom-left corner.

Finally, for all considered values of $K$ and $N$, the patterned domains are clearly below the Curie line at high densities, meaning that the order-by-disorder scenario proposed in 
Ref.~[\onlinecite{Casiulis2019b}] is observed for various values of the density.
In fact, it can even be observed in the phase-separated region for sufficiently low values of $K$ and $N$.

\section{Effect of $\tau$ on the Correlations}
\begin{figure}[b]
    \centering
    \includegraphics[width = .46 \columnwidth]{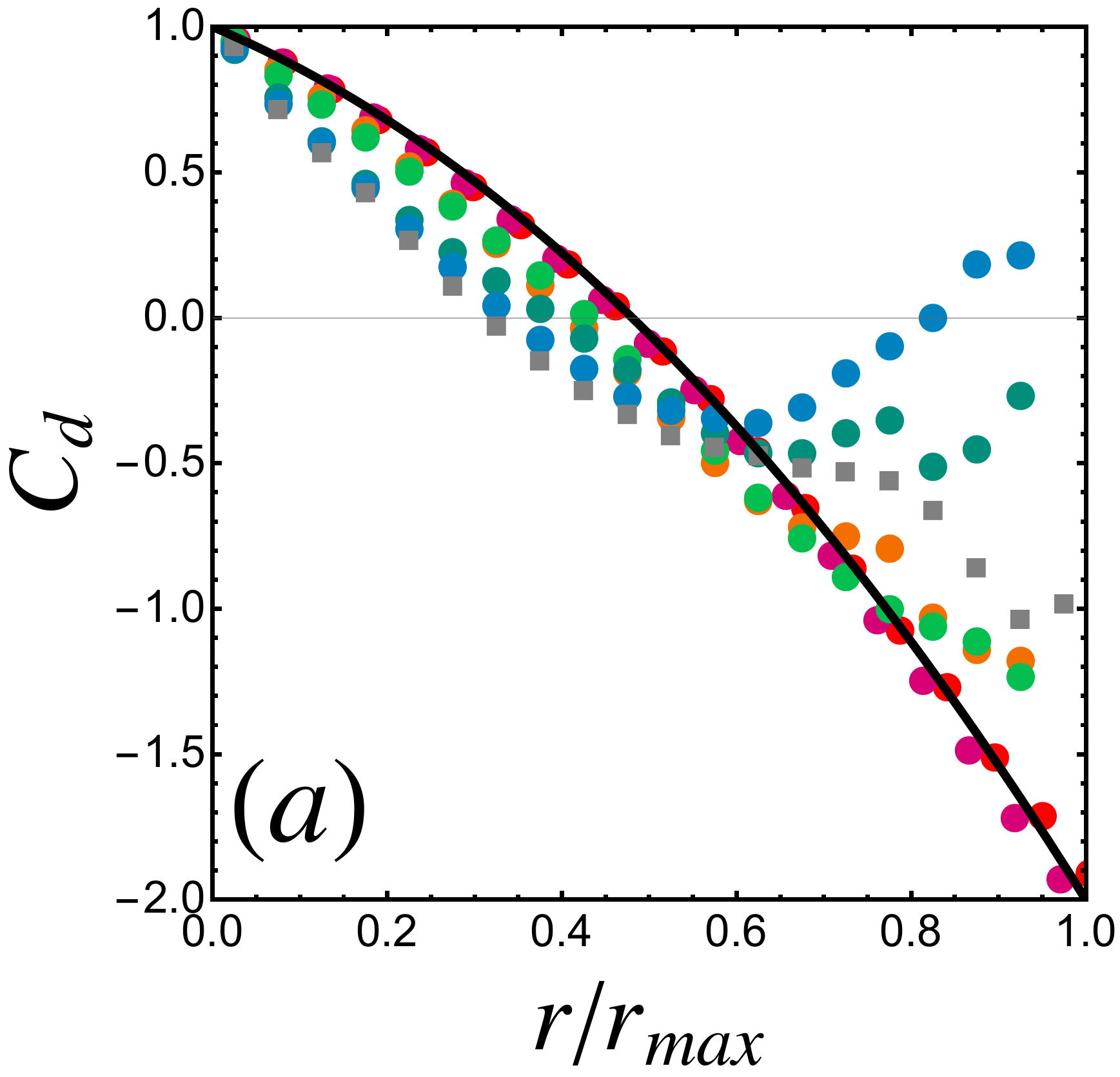}
    \includegraphics[width = .46\columnwidth]{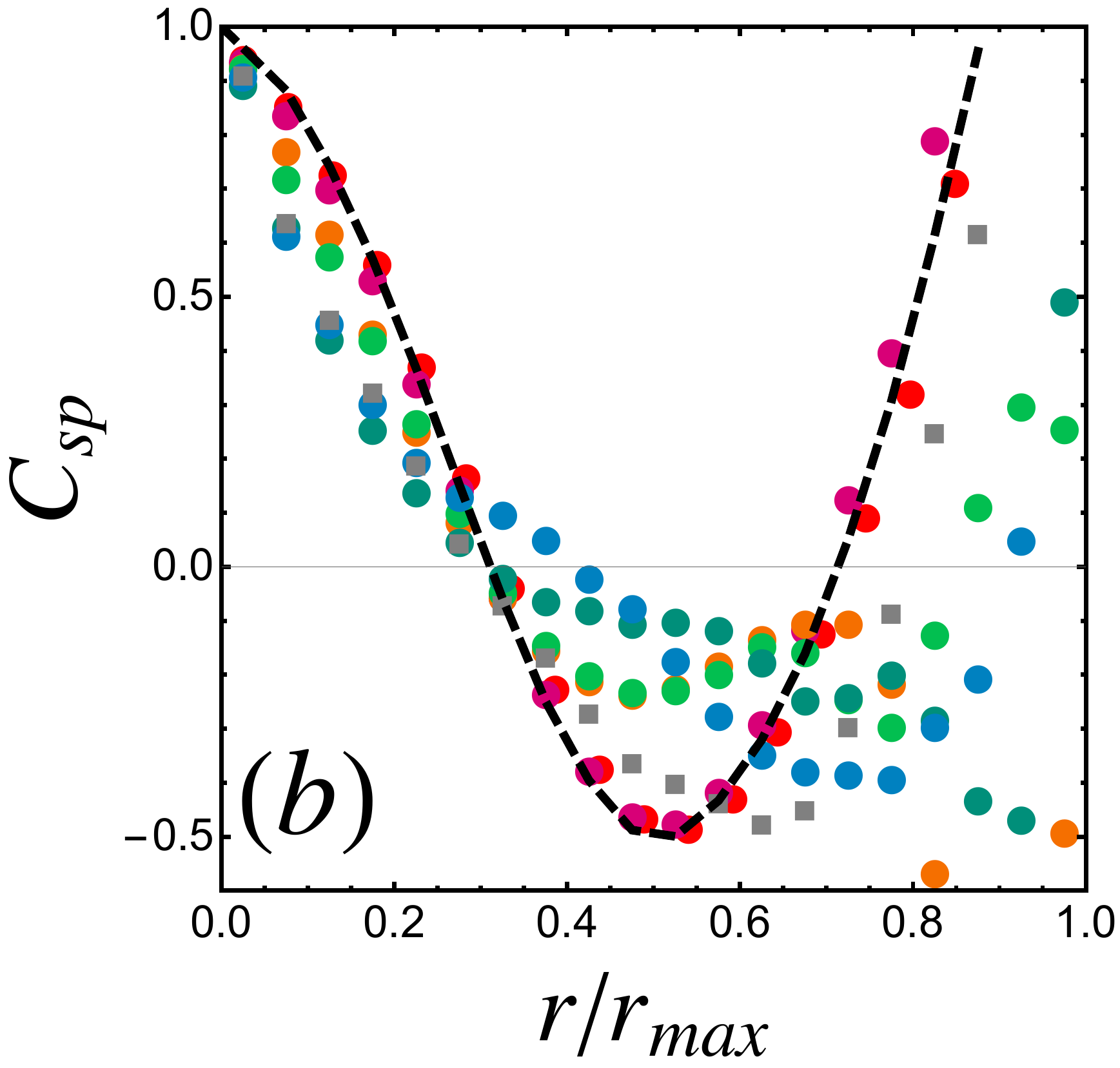}
    \caption{\textbf{Role of the integration time}
    $(a)$ $C_d$ and $(b)$ $C_{sp}$, for $N = 2048, K= 0.001, T \approx 10^{-2}$ (colored disks) and integration times $\tau = 20,30,50,80,250,4000$ growing from blue to red. The black curves are the results for a homogeneous rotating disk (see Sec.~\ref{sec:HomDisk}), 
    and are obtained either with an analytical expression (solid line) or a Monte Carlo integration with $2 \, 10^6$ shots (dashed line). 
    The gray squares were obtained analyzing the data in the SI of Ref.~[\onlinecite{Cavagna2010}].
    }
    \label{fig:IntTime}
\end{figure}
In the main text, when considering displacement-displacement correlations, we choose the integration time $\tau$ long enough 
that we observe the scale-free patterns we describe there.
To show the effect of a short $\tau$ on the correlations of the displacements, we plot in Fig.~\ref{fig:IntTime}$(a)$ (respectively $(b)$) curves of $C_d$ (respectively $C_{sp}$) against the distance normalized by $r_{max}$, for various integration times from $\tau = 20$ (blue) to $\tau = 4000$ (red).
This Hamiltonian flock has $N = 2048$ particles.
As the integration time increases, the correlation functions converge towards their large-time functional form, 
that can be recovered when considering a homogeneous disk that rotates at a constant rate (black lines, see Sec.~\ref{sec:HomDisk}).
The convergence is proved by the fact that the magenta and red points for $\tau=250$ and $\tau=4000$ fall on each other within our numerical accuracy.
In particular, the distance at which the correlation function $C_d$ first cancels increases as $\tau$ is tuned up.
Interpreting this distance as a correlation length, this increase indicates that short-time thermal fluctuations suppress the correlations between displacements.
Regarding $C_{sp}$, small values of $\tau$ also seem to alter the functional form of the correlations at long range, although there is no clear growth of a length.

In order to compare our results to real flocks, we use the data on one flock provided in the SI of Ref.~[\onlinecite{Cavagna2010}].
In this flock, we compute $C_d$ and $C_{sp}$, and plot them in Fig.~\ref{fig:IntTime} using gray squares.
Comparing them to the curves obtained for Hamiltonian flocks, it seems that the functional form of the $C_d$ of Ref.~[\onlinecite{Cavagna2010}] can be reproduced in a Hamiltonian flock with a rather short $\tau$, such that not all fluctuations are smoothed out.
It would be interesting, in real bird flocks, to see whether the relative importance of rotations and deformations (including spin waves) also depend on the chosen integration time.

\section{Effect of $T$ on the Correlations}
\begin{figure}[b]
    \centering
    \includegraphics[width = .46 \columnwidth]{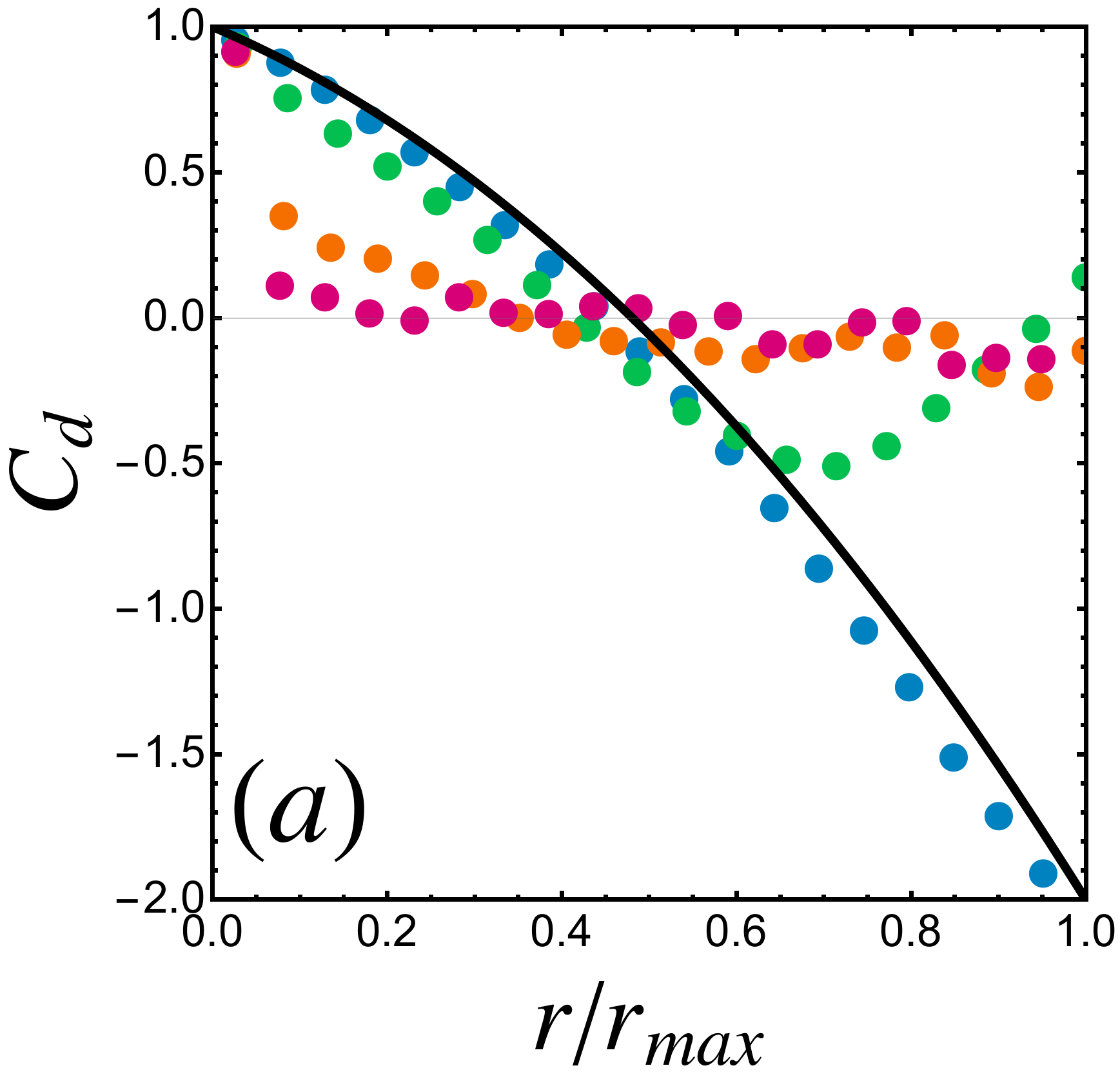}
    \includegraphics[width = .46\columnwidth]{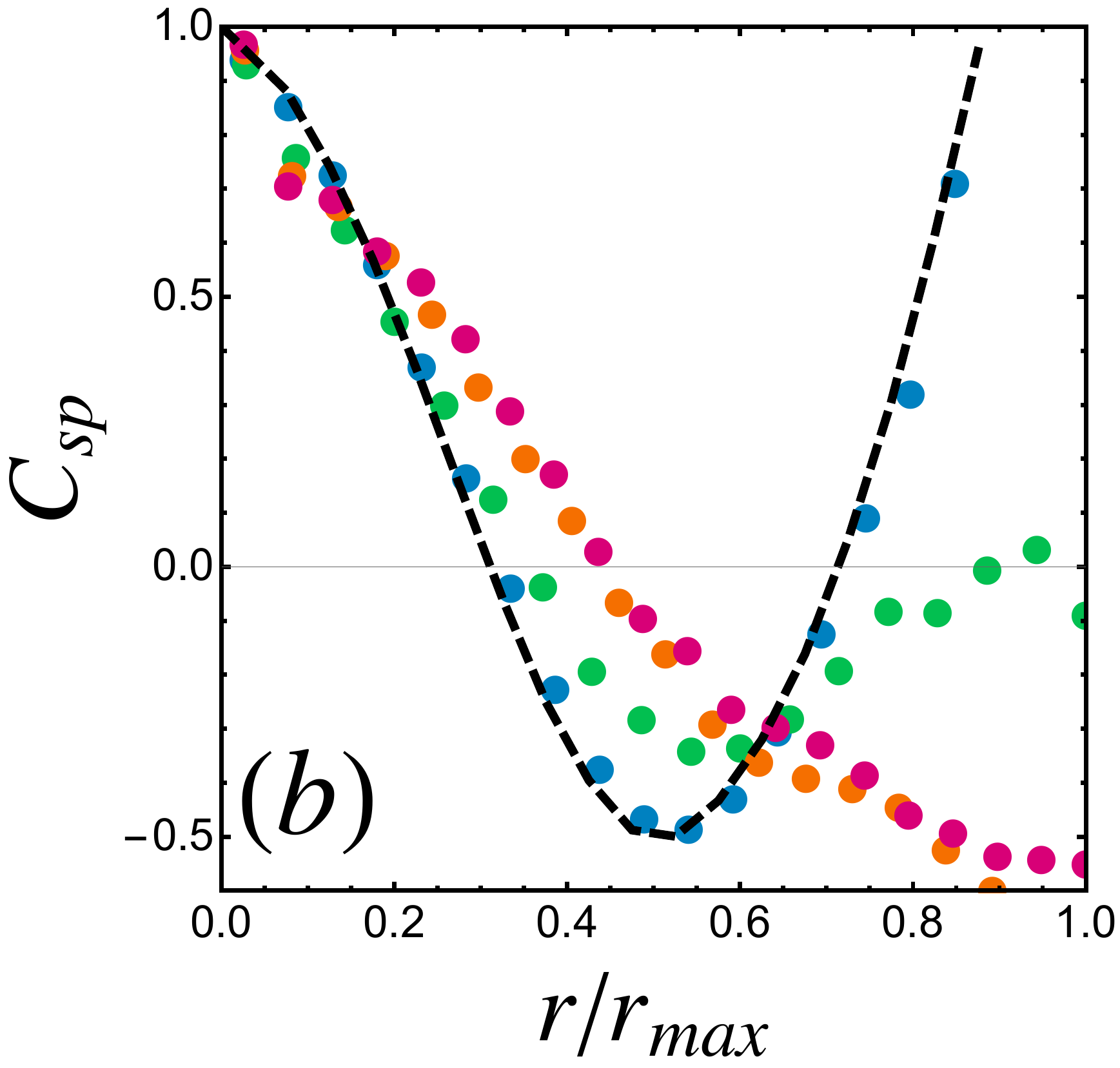}
    \caption{\textbf{Role of the temperature}
    $(a)$ $C_d$ and $(b)$ $C_{sp}$, for $N = 2048, K= 0.001, \tau = 4000$ (colored disks) and temperatures $T = 1.1\cdot 10^{-2},2.0\cdot 10^{-2},2.2\cdot 10^{-2}, 2.4\cdot 10^{-2}$ growing from red to mauve, and crossing the hexatic transition that sits at $T_H \approx 2.0 \cdot 10^{-2}$. The black curves are the results for a homogeneous rotating disk (see Sec.~\ref{sec:HomDisk}), 
    and are obtained either with an analytical expression (solid line) or a Monte Carlo integration with $2 \, 10^6$ shots (dashed line). 
    }
    \label{fig:Temp}
\end{figure}
In the main text, we only focus on the displacement-displacement correlation functions in the rigid regime, as it is the only regime in which the correlation length is the size of the droplet.
In order to illustrate the role played by the temperature on the correlations of the displacements, we show in Fig.~\ref{fig:Temp}$(a)$ (respectively $(b)$) curves of $C_d$ (respectively $C_{sp}$) against the distance normalized by $r_{max}$ for one system size ($N = 2048, K = 0.001, \phi = 0.14$), and a single integration time ($\tau = 4000$), but at various temperatures spanning from $T = 1.1\cdot 10^{-2}$ (blue) to $T =2.4\cdot 10^{-2}$ (red), to be compared with the hexatic transition temperature $T_H \approx 2.0\cdot 10^{-2}$.
These curves show that, as the temperature increases and the system becomes fluid, $C_d$ crosses over from the limit of the rigid disk undergoing a homogeneous rotation to another functional form, that decays faster.
The $C_{sp}$ curves, on the other hand, cross over from the typical U shape of the homogeneous disk to a simple decay.
The precise functional forms of these correlation functions are \textit{a priori} hard to predict, as one would need to formulate the full visco-elasto-plastic theory of cold droplets of this fluid to do it.

\section{Displacement-displacement Correlations in an Homogeneous Disk\label{sec:HomDisk}}

In the main text, we present exact curves for the correlation function $C_d$.
The corresponding calculation was performed in the case of a disk with homogeneous density $\rho_0$ and radius $R$, 
rotating around its center at a constant angular velocity $\Omega$.
In this case, the velocity field can be written in polar coordinates (with the origin at the center of the disk) as
\begin{align}
    \bm{v}(\bm{r}) = r \Omega \, \hat{\bm{e}}_\theta
    \; ,
\end{align}
where $\hat{\bm{e}}_\theta$ is the unit orthoradial vector at position $\bm{r}$.
By construction, this velocity field has a zero sum over the whole disk.
We seek to compute
\begin{align}
    C_d(r) &= \frac{1}{c_0} \frac{\sum\limits_{i\neq j} \bm{v}_i \cdot \bm{v}_j \delta(r_{ij} - r)}{\sum\limits_{i\neq j} \delta(r_{ij} - r)}
    \; ,
\end{align}
where the sum is computed over the particles that constitute the disk, and $c_0$ ensures that $C_d(0) = 1$.
These sums can be approximated by integrals that run over the surface of the disk,
\begin{align}
    C_d(r) &= \frac{1}{c_0} \frac{\int d^2\bm{r}_1 d^2\bm{r}_2 \; \rho(\bm{r}_1) \rho(\bm{r}_2) \bm{v}(\bm{r}_1) \cdot \bm{v}(\bm{r}_2) 
    \, \delta(r_{12} - r)}{\int d^2\bm{r}_1 d^2\bm{r}_2 \; \rho(\bm{r}_1) \rho(\bm{r}_2) \, \delta(r_{12} - r)}
    \; .
\end{align}
Since we here assume the density to be homogeneous, a factor of $\rho_0^2$ comes out of both integrals, and the density disappears from the calculation altogether.
The task at hand is then to compute the integrals in the numerator $\mathcal{N}$ and the denominator $\mathcal{D}$ of $C_d$, defined by 
\begin{align}
    \mathcal{N}(r) &\equiv  \int d^2\bm{r}_1 d^2\bm{r}_2 \; \bm{v}(\bm{r}_1) \cdot \bm{v}(\bm{r}_2) \, \delta(r_{12} - r)
    \; , \\
    \mathcal{D}(r) &\equiv \int d^2\bm{r}_1 d^2\bm{r}_2 \; \delta(r_{12} - r)
    \; .
\end{align}

We first focus on the denominator, $\mathcal{D}$.
The geometric constraint encoded by the Dirac-$\delta$ can be rewritten explicitly in terms of the polar coordinates of the 
positions $\bm{r}_1$ and $\bm{r}_2$, leading to
\begin{align}
        \mathcal{D}(r) &= \int\limits_{0}^R dr_1 \int\limits_{0}^R dr_2 \int\limits_{-\pi}^\pi d\theta_1 \int\limits_{-\pi}^\pi d\theta_2\, r_1 r_2 
        \; \delta\left( \sqrt{r_1^2 + r_2^2 - 2 r_1 r_2 \cos\theta_{12}} - r\right)
        \; ,
\end{align}
where $\theta_{12} = \theta_2 - \theta_1$.
One can then use the change of variables $\theta_2 \to \theta + \theta_1$, integrate over $\theta_1$, and use the periodicity and parity of the cosine to write
\begin{align}
        \mathcal{D}(r) &= 4 \pi \int\limits_{0}^R dr_1 \int\limits_{0}^R dr_2 \int\limits_{0}^\pi  d\theta \; r_1 r_2 \,  \delta\left(\sqrt{r_1^2 + r_2^2 - 2 r_1 r_2 \cos\theta} - r\right). \label{eq:PolarDelta}
\end{align}
The integral over the remaining angle can then be computed, using the fact that for any function $g$ with a single zero at a point $x_0$, one has~\cite{GelFand1968}
\begin{align}
    \int\limits_{a}^{b} dx \; f(x) \, \delta\left(g(x)\right) &= \frac{f(x_0)}{\left|g'(x_0)\right|} \; \mathbb{1}\left( a \leq x_0 \leq b \right),
\end{align}
where $\mathbb{1}$ is a Boolean function which takes the value $1$ if its argument is true, and is $0$ otherwise.
Since the cosine is a bijection from $\left[ 0 ; \pi\right]$ to $\left[ -1 ; 1\right]$, there is \textit{at most} one value $\theta_0$ of $\theta$ that verifies the condition imposed by the $\delta$ in Eq.~(\ref{eq:PolarDelta}).
This value is given by
\begin{align}
        \theta_0 &= \arccos\left[\frac{r_1^2 + r_2^2 - r^2}{2 r_1 r_2}\right],\label{eq:geomcons}
\end{align}
and there is such a value in $\left[0;\pi\right]$ if and only if the argument of the arccos is indeed in $\left[-1;1\right]$, implying the conditions
\begin{align}
(r_1-r_2)^2 \leq r^2 \leq (r_1+r_2)^2
    \; .
\end{align}
Furthermore, we compute the derivative $g'(\theta_0)$ of the argument $g(\theta)$ of the Dirac-$\delta$, 
\begin{align}
    g'(\theta_0) = \frac{r_1 r_2}{r} \sqrt{1 - \left( \frac{r_1^2 + r_2^2 - r^2}{2 r_1 r_2}\right)^2}
    \; .
\end{align}
All in all, integrating over $\theta$ in Eq.~(\ref{eq:PolarDelta}) yields
\begin{align}
        \mathcal{D}(r) &= 4 \pi r \int\limits_{0}^R dr_1  \int\limits_{0}^R dr_2 \; 
        \frac{\mathbb{1}\left[(r_1 - r_2)^2 \leq r^2 \leq (r_1 + r_2)^2 \right]}{\sqrt{1 - \left( \frac{r_1^2 + r_2^2 - r^2}{2 r_1 r_2}\right)^2}}
        \; .
\end{align}
An equivalent expression is obtained by multiplying the numerator and denominator of the integrand by $2 r_1 r_2$,
\begin{align}
        \mathcal{D}(r) &= 8 \pi r \int\limits_{0}^R dr_1\, r_1 \int\limits_{0}^R dr_2\, r_2\frac{\mathbb{1}\left[(r_1 - r_2)^2 \leq r^2 \leq (r_1 + r_2)^2 \right]}{\sqrt{-r_2^4 + 2  r_2^2\left(r_1^2 + r^2\right)- \left( r_1^2 - r^2\right)^2}}
        \; .
\end{align}
We then change variables according to $r_2 \to u = r_2^2$, and notice that the square root in the denominator is a degree-two polynomial in $u$, with roots $u_\pm = \left( r \pm r_1 \right)^2$.
Furthermore, the conditions in the numerator can be expressed as constraints on $u$ rather than on $r$, yielding $u_- \leq u \leq u_+$ (which ensures that the content of the square root is positive), so that
\begin{align}
        \mathcal{D}(r) &= 4 \pi r \int\limits_{0}^R dr_1\, r_1 \int\limits_{0}^{R^2} du \frac{\mathbb{1}\left[u_- \leq u \leq u_+ \right]}{\sqrt{-\left( u - u_+ \right) \left( u - u_- \right)}}
        \; 
        . \label{eq:Denomu}
\end{align}
It is now convenient to use the conditions in the Boolean function to split the integral over $u$ into several domains.
First, since $u_- \geq 0$, the lower bound of the integral over $u$ is in fact always $u_-$.
Furthermore, for the integration domain to have a finite extent, one should have $u_- \leq R^2$, which imposes $r_1 \geq r - R$.
Finally, the upper bound of the integral over $u$, is the lower of the two values $R^2$ and $u_+$.
Stating that $u_+ \leq R^2$ is equivalent to stating that $r_1 \leq R - r$.
Therefore, the $2d$ integration domain naturally splits into two parts: one such that $r_1 \leq R - r$ (provided that $r \leq R$) so that $u$ reaches $u_+$; and one such that $r_1 \geq R - r$ (regardless of the value of $r$) so that $u$ reaches $R^2$ before $u_+$.
In the end, one gets
\begin{align}
            \mathcal{D}(r) &= 4 \pi r \left[\mathbb{1}\left( r \leq R\right)\int\limits_{0}^{R - r} dr_1\, r_1 \int\limits_{u_-}^{u_+} du \frac{1}{\sqrt{-\left( u - u_+ \right) \left( u - u_- \right)}} + \int\limits_{\left|R -r\right|}^{R} dr_1\, r_1 \int\limits_{u_-}^{R^2} du \frac{1}{\sqrt{-\left( u - u_+ \right) \left( u - u_- \right)}}\right]. \label{eq:Denomusplit}
\end{align}

To compute the integrals over $u$, we introduce the change of variables $u = u_- \cos^2\phi + u_+ \sin^2 \phi$.
The corresponding differential elements are related by the equation \begin{align}
    du = 2 \cos\phi \sin\phi \left(u_+ - u_- \right) \, d\phi 
    \; ,
\end{align}
and the polynomial in the square root can be rewritten as
\begin{align}
    - (u-u_+)(u-u_-) &= \left( u_+ - u_- \right)^2 \cos^2\phi \sin^2\phi.
\end{align}
Finally, the bounds of the integral are given by
\begin{align}
    \phi_R &= \arcsin\left[\sqrt{\frac{R^2 - u_-}{u_+ - u_-}} \right], \\
    \phi_- &= 0 \; , \\
    \phi_+ &= \frac{\pi}{2} \; ,
\end{align}
when $u = R^2$, $u = u_-$, and $u = u_+$, respectively.
With all these transformations, Eq.~(\ref{eq:Denomusplit}) can be rewritten as
\begin{align}
            \mathcal{D}(r) &= 8 \pi r \left[\mathbb{1}\left( r \leq R\right)\int\limits_{0}^{R - r} dr_1\, r_1 \int\limits_{0}^{\pi/2} d\phi  + \int\limits_{\left|R -r\right|}^{R} dr_1\, r_1 \int\limits_{0}^{\phi_R} d\phi \right]. \label{eq:Denomusplitphi}
\end{align}
The first term is an elementary integration, while the second one, that we shall call $\mathcal{I}$, is a bit more challenging.
It reads
\begin{align}
    \mathcal{I} &\equiv \int\limits_{|R-r|}^{R} dr_1 \; r_1 \arcsin\left[\sqrt{\frac{R^2 - \left(r - r_1\right)^2}{4 r r_1}} \right].
\end{align}
An integration by parts, as well as some simplifications yield
\begin{align}
    \mathcal{I} &= \left[ \frac{r_1^2}{2} \arcsin\left[\sqrt{\frac{R^2 - \left(r - r_1\right)^2}{4 r r_1}} \right] \right]_{|r - R|}^R - \frac{1}{2} \int\limits_{|R-r|}^{R} dr_1 r_1 \frac{r^2 - R^2 - r_1^2}{\sqrt{-r_1^4 + 2 r_1^2 (R^2 + r^2) - (R^2 - r^2)^2}}
    \; .
\end{align}
The last remaining integral can, like before, be treated by first introducing the variable $v = r_1^2$ and then writing 
$v = v_- \cos^2 \phi + v_+ \sin^2 \phi$, with $v_\pm \equiv (r \pm R)^2$.
Applying these changes one can finally calculate the remaining integral and 
obtain an explicit expression for the denominator $\mathcal{D}$,
\begin{align}
    \mathcal{D}(r) &= \pi r R^2\left(\pi - \frac{r}{R} \sqrt{4-\frac{r^2}{R^2}} + 4\, \text{arccsc}\frac{2}{\sqrt{2 -r/R}} - 2 \arctan\frac{r/R}{\sqrt{4 - r^2/R^2}} \right).
\end{align}
Note that this function is closely related to the pair correlation function in the disk.
The latter is essentially recovered by dividing $\mathcal{D}$ by a factor of $2 \pi r$.

We now turn our attention to the numerator $\mathcal{N}(r)$.
Following the same steps that led to Eq.~(\ref{eq:PolarDelta}) one deduces
\begin{align}
        \mathcal{N}(r) &= 4 \pi \omega^2 \int\limits_{0}^R dr_1 \int\limits_{0}^R dr_2 \int\limits_{0}^\pi  d\theta \; r_1^2 r_2^2 \, \cos\theta \,  \delta\left( \sqrt{r_1^2 - r_2^2 + 2 r_1 r_2 \cos\theta} - r\right). \label{eq:PolarDeltaNum}
\end{align}
The integration over $\theta$ can be treated in the same way as for $\mathcal{D}$, leading to the expression
\begin{align}
        \mathcal{N}(r) &= 2 \pi r \omega^2 \int\limits_{0}^R dr_1  \int\limits_{0}^R dr_2 \; 
        \frac{r_1^2 + r_2^2 - r^2 }{\sqrt{1 - \left( \frac{r_1^2 + r_2^2 - r^2}{2 r_1 r_2}\right)^2}} 
        \; \mathbb{1}\left[(r_1 - r_2)^2 \leq r^2 \leq (r_1 + r_2)^2 \right].
\end{align}
The changes of variables $u = r_2^2$ and $u = u_- \cos^2 \phi + u_+ \sin^2 \phi$, with the same definitions as before, then leads to
\begin{align}
        \mathcal{N}(r) &= 4 \pi r \omega^2 \int\limits_{0}^{R-r} dr_1 r_1 \int\limits_{0}^{\pi/2} d\phi \left[ \left(r_1^2 - r^2\right) + u_- \cos^2\phi + u_+ \sin^2\phi \right] \; \mathbb{1}\left[ r \leq R \right] \nonumber \\
        &\quad + 4 \pi r \omega^2 \int\limits_{\left|R-r\right|}^{R} dr_1 r_1 \int\limits_{0}^{\phi_R} d\phi \; 
        \left[ \left(r_1^2 - r^2\right) + u_- \cos^2\phi + u_+ \sin^2\phi \right].
\end{align}
The first integral is, again, simple to compute.
The second one, that we shall call $\mathcal{J}$, can be rewritten using the expression of $\phi_R$ and some trigonometry ($\sin 2 \theta_R = 2 \cos \theta_R \sin \theta_R$ and $\cos \arcsin x = \sqrt{1 - x^2}$), 
\begin{align}
    \mathcal{J} &\equiv \int\limits_{\left|R-r\right|}^{R} dr_1 r_1 \int\limits_{0}^{\phi_R} d\phi \left[ \left(r_1^2 - r^2\right) + u_- \cos^2\phi + u_+ \sin^2\phi \right] \nonumber \\
    &= 2 \int\limits_{\left|R-r\right|}^{R} dr_1 \; r_1^3  \arcsin\sqrt{\frac{R^2 - \left(r - r_1\right)^2}{4 r r_1}}
     - 2 r \int\limits_{\left|R-r\right|}^{R} dr_1 \; r_1^2 \sqrt{1 - \frac{R^2 - \left(r - r_1\right)^2}{4 r r_1}} \sqrt{\frac{R^2 - \left(r - r_1\right)^2}{4 r r_1}}
    \; .
\end{align}
The first term in $\mathcal{J}$ is similar to $\mathcal{I}$: the only difference is that there is an extra factor of $r_1^2$.
The second term, on the other hand, can be rewritten using the variable $v = r_1^2$, as 
\begin{align}
    \int\limits_{\left|R-r\right|}^{R} dr_1 r_1^2 \sqrt{1 - \frac{R^2 - \left(r - r_1\right)^2}{4 r r_1}} \sqrt{\frac{R^2 - \left(r - r_1\right)^2}{4 r r_1}} &= \frac{1}{8r} \int\limits_{\left(R-r\right)^2}^{R^2} dv \sqrt{- (v - v_+)(v-v_-)}
\end{align}
and can be integrated using techniques similar to those presented above.
The numerator $\mathcal{N}$ can then be written as
\begin{eqnarray}
    \mathcal{N}(r) &=& \frac{\pi \omega^2 R^4}{4} r \left[\pi \left(  2 - 4\frac{r^2}{R^2}\right) + \frac{r}{R} \sqrt{4 - \frac{r^2}{R^2}} \left( \frac{r^2}{R^2} - 6\right) + 16 \frac{r^2}{R^2} \text{arccot}\sqrt{\frac{2 - r/R}{ 2 + r/R}} \right.   \nonumber \\ 
     && +  \left. 8 \text{arccsc}\frac{2}{\sqrt{2 -r/R}} - 4 \left(1 + 2\frac{r^2}{R^2} \right)\arctan \frac{r/R}{\sqrt{4 -r^2/R^2}}\right].
\end{eqnarray}
One can check that $\int_0^{2R} dr \; \mathcal{N}(r) = 0$, which follows from $\sum_{i} \bm{v}_i= \bm{0}$.~\cite{Cavagna2010}

Putting $\mathcal{N}$ and $\mathcal{D}$ back together, we derive an expression for $C_d$.
In order to have $C_d(0) = 1$, one should set $c_0 = R^2 \omega^2 / 4$.
When replacing $c_0$ by this value, the expression for $C_d$ turns out to only depend on $x = r / (2 R)$, and it can be written as
\begin{align}
    C_d(x) &= \frac{\pi \left(  1 - 8 x^2\right) + 4 x \sqrt{1 - x^2} \left( 2x^2 - 3\right) + 32 x^2 \text{arccot}\sqrt{\frac{1 - x}{ 1 + x}}  + 4 \text{arcsin}\sqrt\frac{1 - x}{2} - 2 \left(1 + 8x^2 \right)\arctan \frac{x}{\sqrt{1 - x^2}}}{ 3 \pi - 4x \sqrt{1-x^2} - 4\, \text{arcsec}\frac{2}{\sqrt{2 - 2 x}} - 2 \arctan\frac{x}{\sqrt{1 - x^2}} }
    \; . \label{eq:ExactCd}
\end{align}
As mentioned in the main text, this function only depends on distance through its ratio 
to the size of the system.
Since this function is rather cumbersome, one can also Taylor-expand it around $0$ up to order $2$, finding
\begin{align}
    C_d(x) &\approx 1 - \frac{4}{\pi}x - \frac{16}{\pi^2} x^2
    \; . 
    \label{eq:ApproachedCd}
\end{align}
This expression is a rather good approximation of $C_d$ in the $\left[0;1\right]$ interval.
In particular, it has a root at
\begin{align}
    x_0 = \frac{\pi}{8}\left( \sqrt{5} - 1\right) \approx 0.485
    \; .
\end{align}
\begin{figure}
    \centering
    \includegraphics[width = .45\columnwidth]{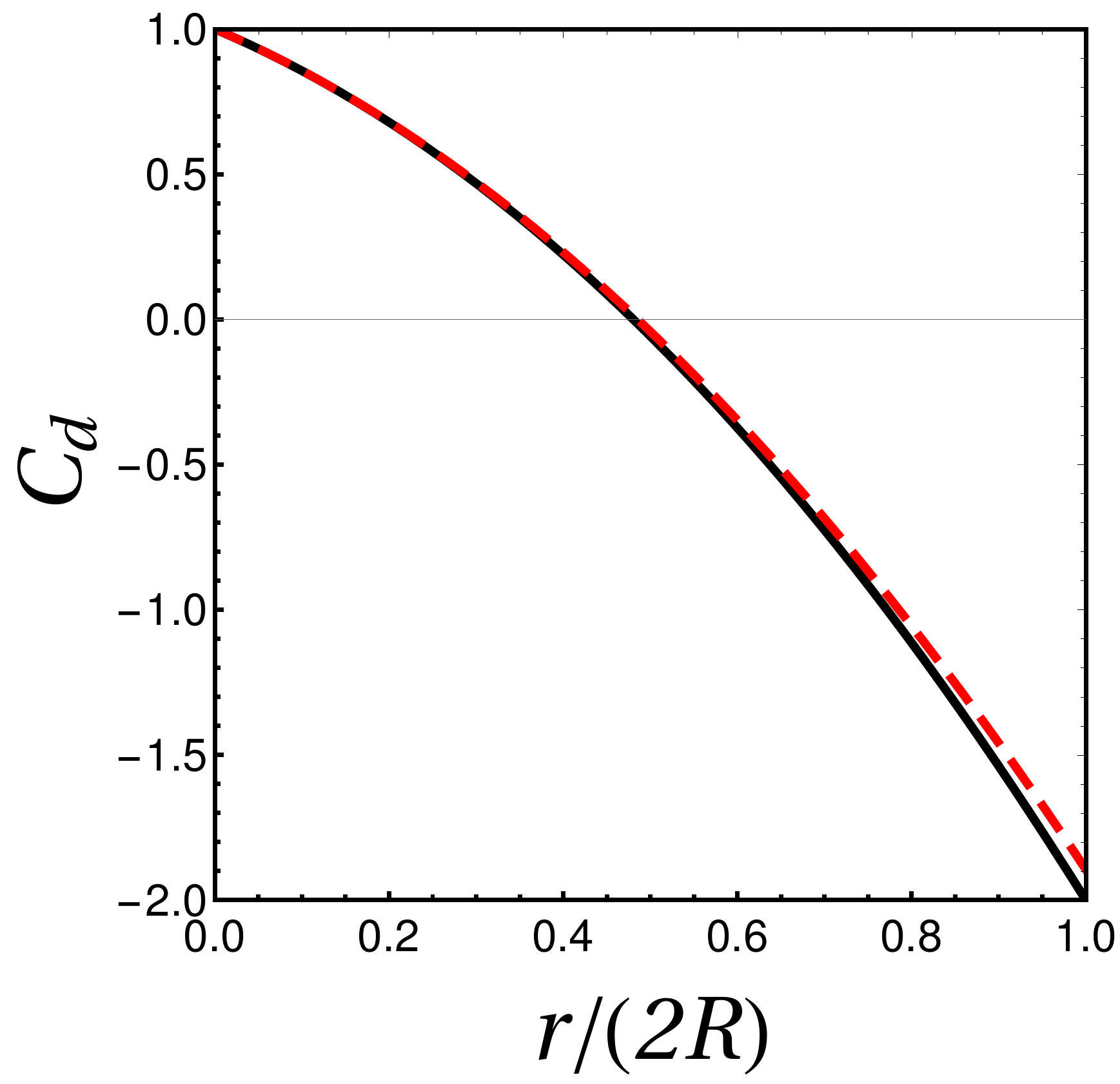}
    \caption{\textbf{Exact displacement-displacement correlations in a rotating homogenous disk.} We show here a plot of the exact expression~(\ref{eq:ExactCd}) (black line), and 
    its Taylor expansion to second order around $0$ (dashed red line), as given by Eq.~(\ref{eq:ApproachedCd}).}
    \label{fig:my_label}
\end{figure}

It is tempting to reproduce the calculation above in the case of the speed-speed correlation $C_{sp}$.
In the case of a homogeneous disk, noticing that 
\begin{align}
    \frac{1}{N} \sum\limits_{i} v_i \approx \frac{1}{\pi R^2} \int\limits d^2\bm{r} \; r \omega = \frac{2}{3}R \omega
    \; , 
\end{align}
the full speed-speed correlation can be written as
\begin{align}
    C_{sp}(r) &= \frac{\omega^2}{c_{0,sp} \mathcal{D}(r)}\int d^2\bm{r}_1 d^2\bm{r}_2 \left(r_1 - \frac{2}{3}R \right)\left(r_2 - \frac{2}{3}R \right) \delta\left( r_{12} - r\right)
     \\
    &=  \frac{\omega^2}{c_{0,sp} \mathcal{D}(r)}\int d^2\bm{r}_1 d^2\bm{r}_2 \left(r_1 r_2 - \frac{4}{3}R r_1 + \frac{4}{9}R^2 \right) \delta\left(r_{12} - r\right)
    \; .
\end{align}
While the last two terms in the integral in the numerator can be computed using the strategies used for $C_d$, the first term is in fact much less tractable.
Indeed, if we call
\begin{align}
    \mathcal{N}_{sp}^1(r) \equiv \omega^2 \int d^2\bm{r}_1 d^2\bm{r}_2 \; r_1 r_2 \, \delta\left(r_{12} - r\right),
\end{align}
following the same route as for $\mathcal{N}$ yields
\begin{align}
    \mathcal{N}_{sp}^1(r) &= 8 \pi r \omega^2 \int\limits_{0}^R dr_1 \; r_1^2 \int\limits_{0}^{R^2} du \; 
    \frac{\sqrt{u}}{\sqrt{- \left( u - u_+\right)\left( u - u_- \right) }} \; \mathbb{1}\left[u_- \leq u \leq u_+ \right].
\end{align}
Here, the most convenient change of variable is the one used for the other integrals, but where $u_+$ and $u_-$ are swapped,
\begin{align}
    u = u_+ \cos^2 \phi + u_- \sin^2 \phi.
\end{align}
Indeed, using this definition, after some more algebra, one finds
\begin{align}
    \mathcal{N}_{sp}^1(r) &= 8 \pi r \omega^2  \int\limits_{0}^{R-r} dr_1 \; r_1^2 \left(r + r_1\right) E\left(\frac{4 r r_1}{\left(r + r_1 \right)^2} \right) \mathbb{1}\left[r \leq R \right]] \nonumber \\
    &\quad + 8 \pi r \omega^2 \int\limits_{\left|r - R \right|}^{R} dr_1 \; r_1^2 \left( r + r_1 \right) \left[ E\left(\frac{4 r r_1}{\left(r + r_1 \right)^2} \right) - E\left(\varphi_R,\frac{4 r r_1}{\left(r + r_1 \right)^2} \right)\right],
\end{align}
where $E(\varphi,x)$ is the incomplete elliptic integral of the second kind,~\cite{Abramowitz1972} $E(x) = E(\pi/2, x)$ is the elliptic integral of the second kind, and
\begin{align}
    \varphi_R &\equiv \arcsin\sqrt{\frac{u_+ - R^2}{u_+ - u_-}}
    \; .
\end{align}
Since elliptic functions are in general hard and cumbersome to treat analytically, there is little hope of writing a tractable expression for $C_{sp}$ using this integration strategy.
However, one still expects an expression that only depends on $r/(2R)$: changing variables following $r_{1,2} \to 2 R x_{1,2}$ in every integral of $C_{sp}$ yields a $R^2 \omega^2$ prefactor that should be absorbed by $c_{0,sp}$, like in the case of $C_d$. 

\section{Rotations in Flocks of Birds}

\begin{figure}[htbp]
    \centering
    \includegraphics[width = 1.0\columnwidth]{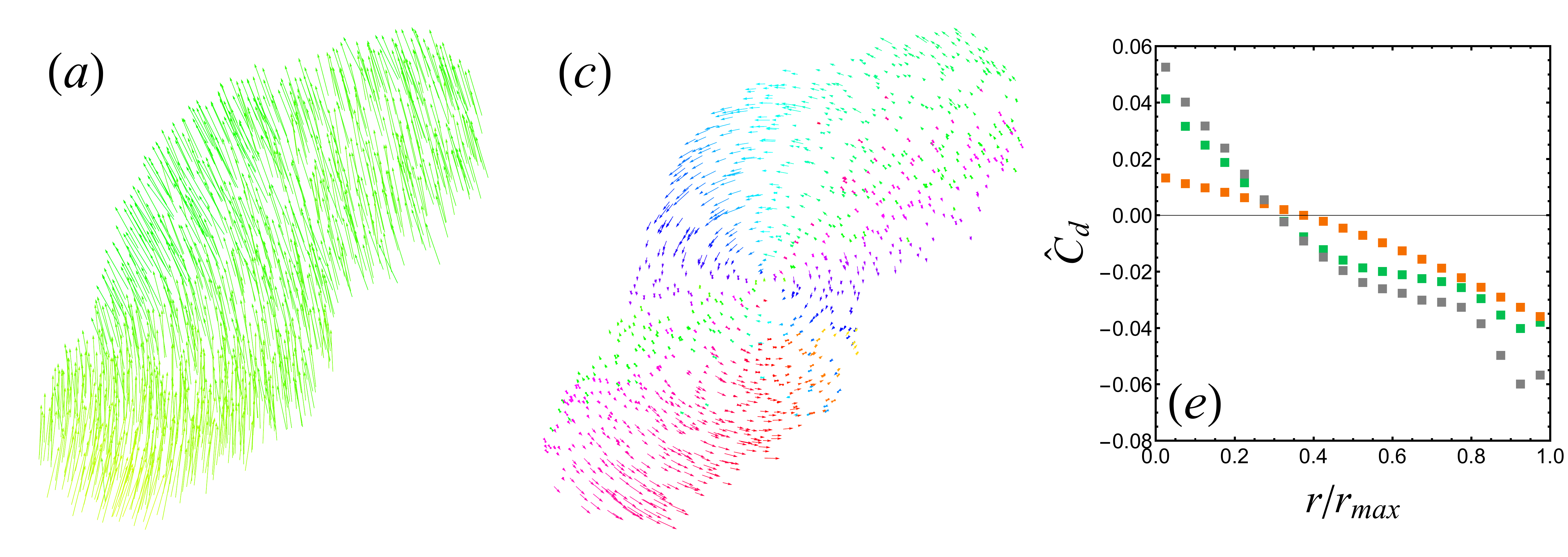} \\
    \includegraphics[width = 1.0\columnwidth]{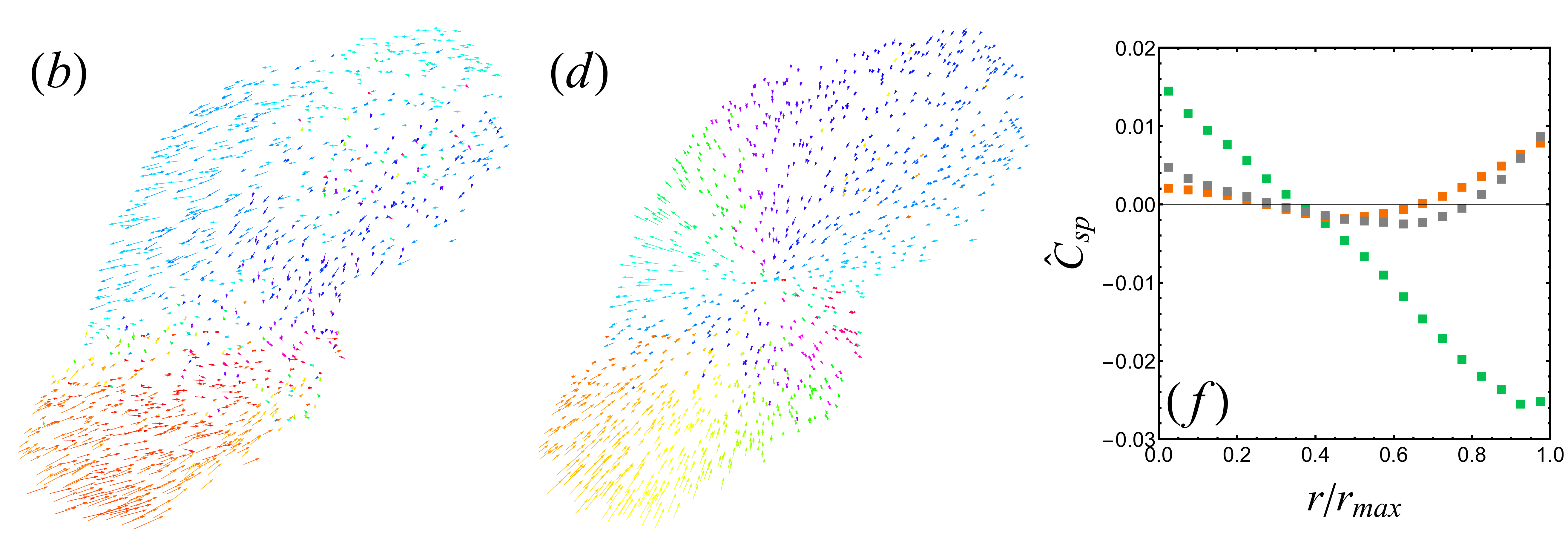}
    \caption{ {\textbf{Bird Displacement Fluctuations.}} 
    $(a)$ Total displacement and $(b)$ displacement fluctuations, as seen from the direction of the main rotation axis.
    $(c)$ Azimuthal and $(d)$ radial components of the displacement fluctuations around the main rotation axis in the flock.
    All displacements are color-coded according to their direction, and multiplied by a constant factor of $3$ for visualization.
    $(e)$ Unnormalized displacement correlations $\hat{C}_d(r)$ against distance normalized by the largest distance in the flock, computed for the full displacements (gray), the pure rigid body rotation (orange), and the rotationless displacement field (green).
    $(f)$ Corresponding curves for $\hat{C}_{sp}$.
    }
    \label{fig:CavagnaSI}
\end{figure}

As shown in the main text, pure rigid body rotations produce displacement-displacement correlations that are very similar to those reported in flocks of birds.\cite{Cavagna2010}
Here, using the data in the SI of Ref.~[\onlinecite{Cavagna2010}], we discuss the magnitude of the correlations induced by rigid body rotations against the correlations induced by other sources of correlations, like Goldstone modes, in one example of a real flock.
In order to do so, we isolate the rigid body rotation part of the displacement field as follows.
We first remove from the total displacement field (Fig.~\ref{fig:CavagnaSI}$(a)$) the translation of the center of mass $\bm{u}_G = N^{-1} \sum \bm{u}_i$ and, thereby, define the relative displacements $\bm{u}_i^\star = \bm{u}_i - \bm{u}_G$, as represented in Fig.~\ref{fig:CavagnaSI}$(b)$.
We next seek the main rigid body rotation axis $(\Delta)$, by computing the sum $\sum\bm{r}_i^\star \times \bm{u}_i^\star$, where $\bm{r}_i^\star = \bm{r}_i - \bm{r}_G$ is the position of bird $i$ relative to the center of mass of the flock.
We then use this axis as the zenith direction of a spherical coordinate system, centered on the centre of mass of the flock, and we decompose the displacement fluctuations into their radial, polar, and azimuthal components.
We find that the polar component $u_{i,\theta}^\star$ (carried by the $\hat{\bm{e}}_{i,\theta}$ unit vector in polar coordinates) is typically much smaller than the other two, so that we omit it in this discussion.
The azimuthal component $u_{i,\phi}^\star$ (carried by the $\hat{\bm{e}}_{i,\phi}$ unit vector in polar coordinates), plotted in Fig.~\ref{fig:CavagnaSI}$(c)$ contains the rotation part of the displacements around the chosen zenith direction, as well as part of the deformations.
The radial component $u_{i,r}^\star$ (carried by the $\hat{\bm{e}}_{i,r}$ unit vector in polar coordinates), plotted in Fig.~\ref{fig:CavagnaSI}$(d)$, contains the radial part of the deformations.
Notice that $u_{i,r}^\star$ roughly assumes the shape of a quadripolar field: the long direction of the flock is compressed, and the short one expanded.
If this shape is a general one across different flocks, it would mean that flocks of birds have a finite Poisson ratio, a property that is reminiscent of elastic solids~\cite{Landau1986}.

In order to isolate the rotation, we fit $u_{i,\phi}^\star (r_i^\star)$ by a linear law, $u_{i,\phi}^\star (r_i^\star) = r_i^\star \Delta \theta$, with $\Delta \theta$ the angle of the rotation.
We find $\Delta \theta \approx 10^{-2} \text{ rad}$ or, equivalently, $\Delta \theta \approx 0.6 \text{ degrees}$.
We then define the rigid body rotation part of the displacement of bird $i$, $\bm{u}_{i}^{rot} = r_i^\star \Delta \theta \hat{\bm{e}}_{i,\phi}$, as well as the rotationless displacement field $\tilde{\bm{u}}_{i} = \bm{u}_i - \bm{u}_{i}^{rot}$.
Finally, we introduce the \textit{unnormalized} correlation functions $\hat{C}_d$ and $\hat{C}_{sp}$, that are defined as in the main text except that we now fix $c_0 = 1$ in both definitions.
These functions are plotted in Fig.~\ref{fig:CavagnaSI}$(e)-(f)$ for the total displacement field $\bm{u}_i$ (gray), the pure rigid body rotation field $\bm{u}_i^{rot}$ (orange), and the rotationless displacement field $\tilde{\bm{u}}_{i}$ (green).
For both $C_d$ and $C_{sp}$, rotations and Goldstone modes yield correlations which only depend on the size of the flock (see Sec.~\ref{sec:HomDisk} and Ref.~\onlinecite{Cavagna2010}).

\newpage
Regarding $\hat{C}_d$, panel $(e)$, we observe that the total correlation is almost identical to the correlation due to the rotationless part of the displacement field.
Consequently, the correlation of the displacement vectors is dominated by deformations.
Since the spin-wave fluctuations of the displacement field are included in deformations, this is compatible with the shape of $C_d$ being dominated by Goldstone modes, as suggested in Ref.~[\onlinecite{Cavagna2010}].
However, the curves of $\hat{C}_{sp}$, panel $(f)$, show that the total correlation is almost identical to the correlation due to the pure rigid rotation alone.
It therefore seems that, in this particular flocking event at least, the shape of the correlation $C_{sp}$ is dictated by a rigid body rotation.

Note that, using the definitions of the main text, $C_d$ and $C_{sp}$ are both \textit{connected} correlation functions: using $\left\langle \cdot \right\rangle_{\bm{r}_0}$ as a short-hand notation for a mean value over the spatial position $\bm{r}_0$, one could write that 
\begin{align}
  \hat{C}_d(r) &= \left\langle \left(\bm{u}(\bm{r_0}) - \left\langle \bm{u} \right\rangle \right)\cdot \left(\bm{u}(\bm{r_0 + \bm{r}}) - \left\langle \bm{u} \right\rangle \right) \right\rangle_{\bm{r}_0}, \\
  \hat{C}_{sp}(r) &= \left\langle \left(\left|\bm{u} \right|(\bm{r_0}) - \left\langle \left|\bm{u} \right| \right\rangle \right)\left(\left|\bm{u} \right|(\bm{r_0 + \bm{r}}) - \left\langle\left|\bm{u} \right| \right\rangle \right) \right\rangle_{\bm{r}_0}.
\end{align}
Several comments on these definitions are in order.
First, notice that in $\hat{C}_d$ the connectedness makes sure that the average translation does not contribute to correlations, so that $\hat{C}_d$ is the same whether you compute it for the total displacement field $\bm{u}$ or the relative displacement field $\bm{u}^\star$.
This is however \textit{not} the case for $\hat{C}_{sp}$: since the modulus is not a linear operation, translations do \textit{a priori} contribute to $\hat{C}_{sp}$, although a pure homogeneous translation $\bm{u} = \bm{cst}$ yields a zero correlation.
Second, one could think of decomposing the displacement field into its rigid part, namely the average translation $\bm{\mathcal{T}}$ and the average rotation around the center of mass $\bm{\mathcal{R}}$ (see Sec.~\ref{sec:kinematics}), and a deformation field $\bm{\mathcal{D}}$.
Using this decomposition, it is straightforward to rewrite $\hat{C}_d$ as
\begin{align}
    \hat{C}_d(r) &= \left\langle \bm{\mathcal{R}}(\bm{r_0}) \cdot\bm{\mathcal{R}}(\bm{r_0}+\bm{r}) \right\rangle_{\bm{r_0}} + \left\langle \bm{\mathcal{D}}(\bm{r_0}) \cdot\bm{\mathcal{D}}(\bm{r_0}+\bm{r}) \right\rangle_{\bm{r_0}} + 2 \left\langle \bm{\mathcal{R}}(\bm{r_0}) \cdot\bm{\mathcal{D}}(\bm{r_0}+\bm{r}) \right\rangle_{\bm{r_0}}.\label{eq:DecompCd}
\end{align}
In this rewriting, the first term corresponds to the orange curve of Fig.~\ref{fig:CavagnaSI}$(e)$, and the second one is the green one.
The fact that the total correlation (gray curve in Fig.~\ref{fig:CavagnaSI}$(e)$) seems to be roughly equal to the sum of the orange and green ones is a sign that the correlation between the rotation and the deformation parts of the displacement field, as defined by the third term in Eq.~(\ref{eq:DecompCd}), is small.
Last but not least, note that this simple decomposition is no longer valid in $\hat{C}_{sp}$ due to the non-linearity introduced by the modulus.
As a result, in Fig.~\ref{fig:CavagnaSI}$(f)$, adding up the green and orange curves has no reason whatsoever to yield the gray curve.

\section{Kinematics of Rigid Turns and Rotations\label{sec:kinematics}}

Finally, to clarify the discussion on rigid body rotations in flocks of birds, we here recall the kinematic definitions of the various kinds of turns and rotations that are usually discussed in bird flocks~\cite{Attanasi2015,Cavagna2018}.
Let us first discuss the cases of \textit{rigid} turns and rotations.
Strictly speaking, rigidity means that the distances between particles remain constant. In practice, there are 
fluctuations and a criterion for rigidity over a chosen time scale $dt$ is that the relative displacements 
satisfy
\begin{equation}
\frac{\Delta | \bm{r}_i - \bm{r}_j| }{|\bm{r}_i-\bm{r}_j|} \ll 1
\qquad\mbox{with}\qquad
\Delta | \bm{r}_i - \bm{r}_j| = | \bm{r}_i - \bm{r}_j|(t+dt) - | \bm{r}_i - \bm{r}_j|(t)
\;,
\label{eq:rigidity}
\end{equation}
where $\Delta | \bm{r}_i - \bm{r}_j|$ is the variation of the length of $\bm{r}_i - \bm{r}_j$ during $dt$.
Let us stress that this definition is purely geometric and does not require nor imply any mechanical property.
We can now give the kinematic definitions (independent of their physical origin) of the terms ``parallel path turn", and ``equal radius turn", that are used in the study of flocks of birds~\cite{Cavagna2018}, and that of a ``rigid body rotation".
\begin{itemize}

\item[-]
A $2d$ rigid set of points, $\bm{r}_i$, performs a parallel path turn when it undergoes a 
pure rotation of an angle $d\theta$ around an arbitrary point $P$
with position $\bm{r}_P$:
\begin{equation}
\bm{r}_i(t+d t) - \bm{r}_i(t) =  [(\bm{r}_i(t) -\bm{r}_P) \times \bm{\Omega}(t) ] dt
\end{equation}
with $\bm{\Omega}(t) dt = d\theta \hat e_z$. 

\item[-]
What is called  a rigid body rotation is a particular case of this motion with $\bm{r}_P = \bm{r}_G$, the position of the
centre of mass $G$. 

\item[-]
In an equal radius turn all points turn around different points, $\bm{r}_{P_i}$, with the same radius of curvature $R$
\begin{equation}
\bm{r}_i(t+d t) - \bm{r}_i(t) =  [(\bm{r}_i(t) -\bm{r}_{P_i}) \times \bm{\Omega}(t) ] dt = \bm{R}(t) \times \bm{\Omega}(t)  dt
\;,
\end{equation}
where $\bm{R} = \bm{r}_i(t) -\bm{r}_{P_i}$ is the same vector with length $R$ for all particles.

\end{itemize}
Any rigid transformation can be uniquely decomposed into a translation and 
a rotation around the centre of mass~\cite{Landau1986}. The parallel path turn contains both, 
the rigid body rotation is a pure rotation around $G$
and the equal radius turn is a pure translation.

In flocks of birds, the picture of a wave-like propagation of a turning information~\cite{Attanasi2014a} implies that birds follow a trajectory akin to an equal radius turn, but with delays between birds.
This transformation is no longer a pure translation and could contain a small amount of rigid body rotation.
Furthermore, it is most likely not a rigid transformation, meaning that it also contains deformations.
All in all, the displacement field of a flock of birds undergoing a turn should, in general, be decomposed as a sum of a translation, a rigid body rotation, and a deformation field.

\bibliography{Bibtex-Droplets-LC}